\RequirePackage{fix-cm}
\documentclass[smallextended]{svjour3}       % onecolumn (second format)
\smartqed  % flush right qed marks, e.g. at end of proof
\usepackage{graphicx}
\usepackage{amsmath,amssymb}
\usepackage{amsfonts}
\usepackage{xspace} 
\usepackage[usenames]{color}
\usepackage{dcolumn}
\usepackage{bm}
\usepackage{mathrsfs}
\usepackage[colorlinks=true]{hyperref}
\usepackage[all]{hypcap} 
\usepackage[utf8]{inputenc}
\usepackage{multirow}
\usepackage{aas_macros} 
\usepackage{hyperref}
\usepackage{cite} % To group references
\newcommand{\eq}{\begin{equation}}
\newcommand{\be}{\begin{equation}}
\newcommand{\eeq}{\end{equation}}
\newcommand{\ee}{\end{equation}}
\newcommand\ba{\begin{eqnarray}}
\newcommand\ea{\end{eqnarray}}
\newcommand{\GB}{{\mbox{\tiny GB}}}
\newcommand{\nn}{\nonumber}
\newcommand{\ppE}{{\mbox{\tiny ppE}}}

\newcommand{\CS}{{\mbox{\tiny CS}}}
\newcommand{\MD}{{\mbox{\tiny MDR}}}

\newcommand{\EDGB}{{\mbox{\tiny EdGB}}}
\newcommand{\dCS}{{\mbox{\tiny dCS}}}

\newcommand{\GR}{{\mbox{\tiny GR}}}
\newcommand{\KG}{{\mbox{\tiny KG}}}

\newcommand{\EH}{{\mbox{\tiny EH}}}
\newcommand{\mrm}{\mathrm}
\newcommand{\DDGB}{{\mbox{\tiny D$^2$GB}}}
\newcommand{\ST}{{\mbox{\tiny ST}}}

\begin{document}

\title{Extreme Gravity Tests with Gravitational Waves from Compact Binary Coalescences:\\(I) Inspiral-Merger
}

\titlerunning{Extreme Gravity Tests with GWs from Compact Binaries: (I) Inspiral-Merger}        % if too long for running head

\author{Emanuele Berti         \and
        Kent Yagi \and 
%        Huan Yang \and 
        Nicol\'as Yunes
        }


\institute{E. Berti \at
Department of Physics and Astronomy, The University of Mississippi, University, MS 38677, USA
           \and
           K. Yagi \at
Department of Physics, University of Virginia, Charlottesville, Virginia 22904, USA
           \and
           N. Yunes \at
eXtreme Gravity Institute, Department of Physics, Montana State University, Bozeman, Montana 59717, USA
}

\date{Received: date / Accepted: date}

\maketitle

\begin{abstract}
  The observation of the inspiral and merger of compact binaries by
  the LIGO/Virgo collaboration ushered in a new era in the study of
  strong-field gravity. We review current and future tests of strong
  gravity and of the Kerr paradigm with gravitational-wave
  interferometers, both within a theory-agnostic framework (the
  parametrized post-Einsteinian formalism) and in the context of
  specific modified theories of gravity (scalar-tensor,
  Einstein-dilaton-Gauss-Bonnet, dynamical Chern-Simons,
  Lorentz-violating, and extra dimensional theories). In this
  contribution we focus on (i) the information carried by the inspiral
  radiation, and (ii) recent progress in numerical simulations of
  compact binary mergers in modified gravity.  \keywords{Modified
    Gravity \and Gravitational Waves \and Compact Binary Systems}
\end{abstract}

\tableofcontents

%%%%%%%%%%%%%%%%%%%%%%%%%%%%%%%
\section{Introduction}
\label{intro}

Why would you modify Einstein's beautiful theory of gravity, General Relativity (GR)? This question is asked often in the context of experimental relativity. The driving force behind modified gravity is not to discard Einstein's theory, but rather to address some of the unsolved physics problems that Einstein's theory has brought to the forefront. The late-time expansion of the Universe and dark energy, the rotation curves of galaxies and dark matter, the matter-antimatter asymmetry in the early Universe, the information loss problem and the existence of singularities, the exponential expansion of the early Universe and inflation are just a few of these problems. One could of course assert that these problems are resolved by (highly fine-tuned) initial conditions or the existence of a new form of some yet-undetected weakly interacting particle. Another avenue is to postulate that they are resolved because Nature is better described by a modified gravity theory that reduces to Einstein's in some well-studied scenarios (like in the Solar System or in binary pulsars), but that introduces key modifications in other scenarios (like on large scales, in the early Universe or near black holes). 

One way to classify these theory-inspired modifications to gravity is through the fundamental pillars of GR that they violate or deform. Einstein's theory can be thought of as descending from three organizing principles: diffeomorphism invariance, the (strong) equivalence principle, and spacetime as a four-dimensional continuum. Indeed, one can argue that these principles essentially require spacetime to be curved, and that the need for the theory to reduce to Newtonian gravity in the weak-field limit and the Bianchi identities then lead to the Einstein equations. The resulting theory is necessarily parity invariant (on any spacelike hypersurface) and gravity is described by the curvature of a rank-2 tensor field that is free to propagate at the speed of light, i.e.~in a quantized picture it would be described by a massless, spin-2 particle. Given these pillars, one can imagine breaking or deforming them to address some of the unsolved problems mentioned above. For example, the observed matter-antimatter asymmetry that develops in the radiation-dominated era requires additional sources of parity violation by the Sakharov conditions~\cite{Sakharov:1967dj}, which could be addressed by introducing gravitational parity-violating interactions in the Einstein-Hilbert action~\cite{Alexander:2004xd,Alexander:2009tp}. Another example is the late-time acceleration of the Universe, which instead of being described by a cosmological constant may perhaps be explained by an additional tensor field that interacts with the gravitational field, as in bi-gravity theories~\cite{Pilo:2011zz,Paulos:2012xe}.    

Given so many modified gravity alternatives that have been postulated to address these unsolved problems, how do we decide which one, if any, is the best description of Nature? Experiments and observations are the fool-proof method of choice. Many of these modified theories can be straightforwardly ruled out with current Solar System~\cite{Will:2014kxa}, binary pulsar~\cite{Stairs:2003eg} and cosmological observations. For example, the Brans-Dicke-Jordan-Fierz flavor of scalar-tensor theories~\cite{Brans:1961sx,Fierz:1956zz,Jordan:1959eg} is incompatible with Shapiro time-delay observations by the Cassini spacecraft~\cite{Bertotti:2003rm}, unless the coupling constant is tuned to be sufficiently small. These observations, together with the cosmological evolution of scalar fields in the scalar-tensor theories studied by Damour and Esposito-Far\`ese~\cite{Damour:1992we,Damour:1993hw,Damour:1996ke,Damour:1992kf,Damour:1993id},
prohibit the spontaneous scalarization of neutron stars~\cite{Sampson:2014qqa,Anderson:2016aoi}, unless one fine tunes the cosmological initial conditions of the evolution of the scalar field. 

But even after imposing the requirement that modified theories must pass all current experimental tests, there still remain a large group of them that is only weakly constrained today. One example are effective field theories of gravity that introduce a scalar field sourced by singularities in the spacetime curvature and not by matter~\cite{Yunes:2009hc,Alexander:2009tp,Yunes:2011we}. Such theories will pass Solar System tests because of the weak curvature of spacetime in the Solar System, which thus prevents the scalar field from being strongly excited~\cite{AliHaimoud:2011bk,AliHaimoud:2011fw}. These theories will also pass binary pulsar tests because neutron stars do not possess singularities, and thus, the scalar field is not strongly activated~\cite{Yagi:2011xp,Yagi:2015oca}. One must then rely on new observations, observations that sample the \emph{extreme gravity} regime, where the spacetime is strongly curved and highly dynamical. One example are gravitational wave (GW) observations of the coalescence of binary compact objects, like black holes and neutron stars.  

GW tests are quite different from other tests of gravity that have been carried out to date. As argued above, GWs are unique probes of the extreme gravity regime, being sensitive both to the propagation and the generation of these waves. Moreover, GWs are weakly interacting, and thus, they travel essentially unimpeded to our detectors on Earth, without being affected by intervening matter. For example, the accretion disk of black holes introduces a very small modification to the GWs emitted when black holes collide, except perhaps for the most massive disks~\cite{Kocsis:2011dr,Yunes:2011ws,Hayasaki:2012qn,Barausse:2014tra}. Finally, GW tests are localized in spacetime, sampling the gravitational interaction everywhere inside the wave's lightcone. One could thus imagine that as more GW observations are made, one can begin to build \emph{constraint maps} that represent the verification of the pillars of GR everywhere in the sky. That is, one could envision a GW sky map (similar to that which depicts anisotropies in the cosmic microwave background) that depicts the sky locations of thousands of GW detections and the constraints placed on a given deformation of GR at that particular location, perhaps even allowing constraint contours in sky location.   

The era of GW physics has begun, heralded by the first detections of Advanced LIGO (aLIGO) and Virgo~\cite{Abbott:2016blz,Abbott:2016nmj,Abbott:2017vtc,Abbott:2017oio,TheLIGOScientific:2017qsa,Abbott:2017gyy}. These observations have already confirmed that indeed GWs exist, that they are predominantly quadrupolar in nature, and that they propagate at essentially the speed of light in a Lorentz-invariant fashion. Such tests therefore confirm some of the pillars of GR, but the era of precision experimental relativity with GWs is only beginning. In the next few years, the LIGO/Virgo collaboration will continue to make detections at an ever increasing rate with larger and larger signal-to-noise ratios. 
In the near future new detectors (such as KAGRA in Japan and LIGO-India in India) will join the network, which will allow for precise probes of the polarization content of GWs. Work has already begun to consider upgrades to third-generation detectors, and in the early 2030s space-based detectors will be launched~\cite{Audley:2017drz}. At this stage, the stacking of hundreds or thousands of GW observations will yield unprecedented tests of Einstein's theory of gravity in the mostly unexplored extreme gravity regime.  

This contribution is organized as follows. In Section~\ref{sec:modgrav} we briefly review models of modified gravity that have been explored in some detail in the context of strong-gravity tests, including scalar-tensor theories, Einstein-dilaton Gauss-Bonnet gravity, dynamical Chern-Simons gravity, Lorentz-violating theories, and theories involving extra dimensions. In Section~\ref{sec:inspiral} we present generic and theory-specific tests of modified gravity with compact binary inspirals, reviewing first current constraints, and then future prospects. In Section~\ref{sec:merger} we discuss the current state of numerical simulations of compact binary mergers in modified gravity. We conclude in Section~\ref{sec:outlook} with a list of important open problems. In Appendix~\ref{app:scalar-charge} we derive the scalar charge in decoupled dynamical Gauss-Bonnet gravity for black holes with arbitrary rotation. The final expression was shown in~\cite{Yunes:2016jcc}, but the details of the derivation (and the expression for the quadrupole scalar charge) are presented here for the first time. In Appendix~\ref{app:pn_st} we briefly review a ``dynamical no-hair theorem'' for black holes in scalar-tensor gravity.

%%%%%%%%%%%%%%%%%%%%%%%%%%%%%%%
\section{Modified Theories of Gravity}
\label{sec:modgrav}

%------------------------------------------------------------------
\subsection{Scalar-tensor Theories}
\label{sec:ST}

One of the simplest extensions of GR is {\it scalar-tensor gravity}. In this class of theories, one or more scalar degrees of freedom are included in the gravitational sector through a {\it non-minimal coupling}: the Ricci scalar in the Einstein-Hilbert action is multiplied by a function of the scalar field(s). These theories are well motivated: scalar fields with non-minimal couplings to gravity appear (e.g.) in string theory~\cite{Polchinski:1998rq}, in Kaluza-Klein-like theories~\cite{Duff:1994tn} and in braneworld scenarios~\cite{Randall:1999ee,Randall:1999vf}. These models have also been extensively discussed in a cosmological context~\cite{Clifton:2011jh}. Due to their simplicity, scalar-tensor theories are a good framework to study the strong-field dynamics of modified gravity.  There are many extensive reviews on the subject~\cite{Damour:1992we,Chiba:1997ms,2003sttg.book.....F,Faraoni:2004pi,Sotiriou:2015lxa,Berti:2015itd}, so we will only sketch the main ingredients of these theories -- first in their original form, and then considering generalizations that have attracted some attention in the past few years.

%%%%%%%%%%%%%%%%%%%%%%%%%%%%%%%%%%%%%%%%%%%%%%%%%%%%
\subsubsection{``Bergmann-Wagoner'' scalar-tensor theories}
\label{sec:BergmannWagoner}
%%%%%%%%%%%%%%%%%%%%%%%%%%%%%%%%%%%%%%%%%%%%%%%%%%%%
The most general action of scalar-tensor theory with one scalar field
which is at most quadratic in derivatives of the fields was studied by
Bergmann and Wagoner~\cite{Bergmann:1968ve,Wagoner:1970vr}. With an
appropriate field redefinition, the so-called {\it Jordan-frame}
action for the theory can be cast in the form:
\begin{equation}
  S=\frac{1}{16\pi}\int d^4x\sqrt{-g}\left[\phi R-\frac{\omega(\phi)}{\phi}g^{\mu\nu}
    \partial_\mu\phi\, \partial_\nu\phi
    -U(\phi)\right]+S_M[\Psi,g_{\mu\nu}]\,,\label{STactionJ}
\end{equation}
where $\omega$ and $U$ are arbitrary functions of the scalar field
$\phi$, and $S_M$ is the action of the matter fields $\Psi$. When
$\omega(\phi)=\omega_{BD}$ is constant and $U(\phi)=0$, the theory
reduces to (Jordan-Fierz-)Brans-Dicke
gravity~\cite{Jordan:1959eg,Fierz:1956zz,Brans:1961sx}.

The Bergmann-Wagoner theory \eqref{STactionJ} can be expressed in a
different form through a scalar field redefinition
$\varphi=\varphi(\phi)$ and a conformal transformation of the metric
$g_{\mu\nu}\rightarrow g^\star_{\mu\nu}=A^{-2}(\varphi)
g_{\mu\nu}$. In particular, fixing $A(\varphi)=\phi^{-1/2}$, the
action~(\ref{STactionJ}) transforms into the {\it Einstein-frame}
action\footnote{A straightforward generalization consists of coupling
  gravity with more than one scalar field. Then the
  action~\eqref{STactionJ} has the more general
  form~\cite{Damour:1992we}
\begin{align}
\nonumber
S\,=\,\,& \frac1{16\pi}\int d^4x\sqrt{-g}\left(
F(\phi)R-\gamma_{ab}(\phi)g^{\mu\nu}\partial_\mu\phi^a
\partial_\nu\phi^b-V(\phi)\right)+S_M[\Psi,\,g_{\mu\nu}]\,,
\label{STactionJ_multi}
\end{align}
where $F,V$ are functions of the $N$ scalar fields $\phi^a$
($a=1\ldots N$). The scalar fields live on a manifold (sometimes
called the {\it target space}) with metric $\gamma_{ab}(\phi)$. This
action is invariant not only under space-time diffeomorphisms, but
also under target-space diffeomorphisms, i.e.~scalar field
redefinitions. The geometry of the target space can affect the
dynamics and the structure of compact
objects~\cite{Horbatsch:2015bua}.}
\begin{equation}
  S=\frac{1}{16\pi}\int d^4x\sqrt{-g^\star}\left[R^\star-2g^{\star\mu\nu}
    \left(\partial_\mu\varphi\right)\left(\partial_\nu\varphi\right)
    -V(\varphi)\right]+S_M[\Psi,A^2(\varphi)g^\star_{\mu\nu}]\,,
  \label{STactionE}
\end{equation}
where $g^\star$ and $R^\star$ are the determinant and Ricci scalar of
$g_{\mu\nu}^{\star}$, respectively, and the potential
$V(\varphi) \equiv A^4(\varphi)U(\phi(\varphi))$.  In the Einstein
frame the scalar field is minimally coupled in the gravitational
sector, but there is non-minimal coupling in the matter sector of the
action: particle masses and fundamental constants depend on the scalar
field.

The actions~\eqref{STactionJ} and \eqref{STactionE} are different
representations of the same
theory~\cite{Flanagan:2004bz,Sotiriou:2007zu}, and the frame choice is
often dictated by computational convenience: for instance, in vacuum
the Einstein-frame action \eqref{STactionE} formally reduces to GR
with a minimally coupled scalar field.  However, since the scalar
field is minimally coupled to matter in the Jordan frame, test
particles follow geodesics of the Jordan-frame metric.

Jordan-frame and Einstein-frame quantities are related by
$\phi=A^{-2}(\varphi)$, $3+2\omega(\phi)=\alpha(\varphi)^{-2}$, where
$\alpha(\varphi)\equiv d(\ln A(\varphi))/d\varphi$.
The theory is fixed by choosing the function $\omega(\phi)$ -- or,
equivalently, $\alpha(\varphi)$ -- and the form of the scalar
potential. Neglecting the scalar potential corresponds to neglecting
the cosmological term, the mass of the scalar field and any possible
scalar self-interaction. In an asymptotically flat spacetime the
scalar field tends to a constant $\phi_0$ at spatial infinity,
corresponding to a minimum of the potential. Taylor expanding
$U(\phi)$ around $\phi_0$ yields, to lowest orders, a cosmological
constant and a mass term for the scalar
field~\cite{Wagoner:1970vr,Alsing:2011er}.

Scalar-tensor theory with a vanishing scalar potential is
characterized by a single function $\alpha(\varphi)$. The expansion of
this function around the asymptotic value $\varphi_0$ can be written
in the form
\begin{equation}
\alpha(\varphi)=\alpha_0+\beta_0(\varphi-\varphi_0)+\dots\label{DEalphabeta}
\end{equation}
As mentioned above, the choice $\alpha(\varphi)=\alpha_0=$~constant
(i.e., $\omega(\phi)=$~constant) corresponds to Brans-Dicke theory. A
more general formulation, proposed by Damour and Esposito-Far\`ese, is
parametrized by $\alpha_0$ and
$\beta_0$~\cite{Damour:1993hw,Damour:1996ke}. Another simple variant
is massive Brans-Dicke theory, in which $\alpha(\varphi)$ is constant,
but the potential is non-vanishing and has the form
$U(\phi)=\frac{1}{2}U''(\phi_0)(\phi-\phi_0)^2$, so that the scalar
field has a mass $m_s^2\sim U''(\phi_0)$.

In the Einstein frame, the field equations are
\begin{subequations}
\label{ST:Einstein}  
\begin{align}
  G^\star_{\mu \nu} &=2\left(\partial_\mu\varphi\partial_\nu\varphi-
  \frac{1}{2}g^\star_{\mu\nu}\partial_\sigma\varphi\partial^\sigma\varphi\right)-
  \frac{1}{2}g^\star_{\mu\nu}V(\varphi)+8\pi T^\star_{\mu\nu}\,,\label{eq:tensoreqnE}  \\
\Box_{g^\star} \varphi &=-4\pi\alpha(\varphi)T^\star+\frac{1}{4}\frac{dV}{d\varphi}\,,\label{eq:scalareqnE}
\end{align}
\end{subequations}
where
$T^{\star}_{\mu\nu}$
is the Einstein-frame stress-energy tensor of matter fields and
$T^\star$ is its trace~\cite{Damour:1992we}.
Equation~\eqref{eq:scalareqnE} shows that
$\alpha(\varphi)$ determines the strength of the coupling of the
scalar field to matter~\cite{Damour:1995kt}.

Astrophysical observations set bounds on the parameter space of
scalar-tensor theories. In the case of Brans-Dicke theory, the best
observational bound ($\alpha_0<3.5\times 10^{-3}$) comes from the
Cassini measurement of the Shapiro time delay. In the more general
case with $\beta_0\neq0$, current constraints on $(\alpha_0,\beta_0)$
have been obtained by observations of binary neutron star and neutron star-white dwarf binary
systems~\cite{Freire:2012mg,Shao:2017gwu}, as well as from observations of
the Shapiro time delay upon accounting for the cosmological evolution
of the scalar field~\cite{Sampson:2014qqa,Anderson:2016aoi}.
%, and will be discussed in Section~\ref{sec:BP}
%(cf.~Figure~\ref{fig:stg}).
Observations of compact binary systems also constrain massive
Brans-Dicke theory, leading to exclusion regions in the
$(\alpha_0,m_s)$ plane~\cite{Alsing:2011er}.

%%%%%%%%%%%%%%%%%%%%%%%%%%%%%%%%%%%%%%%%%%%%%%
\subsubsection{Horndeski gravity}\label{subsec:horndeski}
%%%%%%%%%%%%%%%%%%%%%%%%%%%%%%%%%%%%%%%%%%%%%

The most general scalar-tensor theory with second-order field
equations (and one scalar field) is Horndeski
gravity~\cite{Horndeski:1974wa}.  The action of Horndeski gravity can
be written in terms of Galileon interactions~\cite{Deffayet:2011gz} as
\begin{equation}
\begin{aligned}
  S={}&\int d^4x\sqrt{-g}\Big\{K(\phi,X)-G_3(\phi,X)\Box\phi\\
    &{}+G_4(\phi,X)R+G_{4,X}(\phi,X)\left[(\Box\phi)^2
      -(\nabla_\mu\nabla_\nu\phi)(\nabla^\mu\nabla^\nu\phi)\right]\\
    &{}+G_5(\phi,X)G_{\mu\nu}\nabla^\mu\nabla^\nu\phi
    -\frac{G_{5,X}(\phi,X)}{6}\big[
      (\Box\phi)^3-3\Box\phi(\nabla_\mu\nabla_\nu\phi)(\nabla^\mu\nabla^\nu\phi)
    \\
    &{}+2(\nabla_\mu\nabla_\nu\phi)(\nabla^\mu\nabla_\sigma\phi)
    (\nabla^\nu\nabla^\sigma\phi)\big]\Big\}\,,
\end{aligned}
    \label{action_horndeski}
\end{equation}
where $K$ and the $G_i$'s ($i=1\dots 5$) are functions of the scalar
field $\phi$ and of its kinetic term
$X=-(\partial^\mu\phi\partial_\mu\phi)/2$, and $G_{i,X}$ are
derivatives of $G_i$ with respect to the kinetic term $X$.
For a particular choice of these
functions~\cite{Kobayashi:2011nu,Maselli:2015yva} this theory
coincides with Gauss-Bonnet gravity, which will be discussed next.

It was recently understood that Horndeski gravity is a subclass of all
higher-order scalar-tensor theories that contain a single scalar mode.
The crucial ingredient that singles out higher-order theories with a
single scalar degree of freedom is the degeneracy of their Lagrangian,
and now these theories are commonly referred to as Degenerate
Higher-Order Scalar-Tensor (DHOST)
theories~\cite{Langlois:2015skt,Langlois:2017mdk}. Theories similar to
Horndeski gravity have also been constructed for vector (Proca)
fields~\cite{Heisenberg:2014rta,Heisenberg:2016eld}.

%------------------------------------------------------------------
\subsection{Einstein-dilaton Gauss-Bonnet Gravity}
%[Kent]

Einstein-dilaton Gauss-Bonnet (EdGB) gravity is a string-inspired theory that acquires a quadratic-curvature correction to the Einstein-Hilbert action $S_\EH$. Motivated from heterotic superstring theory, the action for EdGB gravity is given by~\cite{Maeda:2009uy}
\be
S = S_\EH + S_\EDGB + S_\phi + S_\mrm{mat}\,,
\ee
where $S_\mrm{mat}$ is the matter action while $S_\EDGB$ and $S_\phi$ are the quadratic-curvature correction and the kinetic term for the scalar field $\phi$, respectively, given by
\be
S_\EDGB  = \alpha_\EDGB \,  \int d^4x \sqrt{-g} e^{-\gamma_\EDGB \phi} R_\GB^2\,, \quad
S_\phi = - \frac{1}{2} \int d^4 x \sqrt{-g} \nabla_\mu \phi \nabla^\mu \phi\,.
\ee
Here $g$ is the determinant of the metric, $\alpha_\EDGB$ and $\gamma_\EDGB$ are coupling constants while 
\be
R_\GB^2 \equiv R^2 - 4 R_{\mu\nu} R^{\mu\nu} + R_{\mu\nu\rho\sigma}R^{\mu\nu\rho\sigma}\,,
\ee
with $R$, $R_{\mu\nu}$ and $R_{\mu\nu\rho\sigma}$ representing the Ricci scalar, Ricci tensor and Riemann tensor, respectively. This specific combination of curvature quantities ensures derivatives in the field equations to be at most of second order, but the well-posedness of this theory is still an open question. We assume that the scalar field is dimensionless in geometrical units, so $\alpha_\EDGB$ has units of length squared, while $\gamma_\EDGB$ is dimensionless. For simplicity, we set the scalar field potential to zero. Current constraints on $\alpha_\EDGB$ come from the orbital decay rate of low-mass X-ray binaries with a black hole~\cite{kent-LMXB}, the existence of stellar-mass black holes~\cite{Pani:2009wy} and the maximum mass of neutron stars~\cite{Pani:2011xm}. Black hole solutions have been constructed numerically for static configurations~\cite{Kanti:1995vq,Torii:1996yi,Alexeev:1996vs} and rotating configurations~\cite{Pani:2009wy,Kleihaus:2011tg,Kleihaus:2014lba} (see also~\cite{Kokkotas:2017ymc} for approximate analytic solutions for static black holes).

One can treat this theory as an effective field theory since we neglect terms higher than cubic order in curvature -- or terms of $\mathcal{O}(\alpha_\EDGB^2)$ -- in the action. Namely, one can assume that $S_\EDGB$ is much smaller than $S_\EH$. This can be done by perturbing the action about $\phi = 0$ and retaining terms up to linear order. Since $R_\GB^{2}$ can be written in a total derivative form upon variation of the action, the leading term where $R_\GB^{2}$ does not couple to the scalar field does not contribute to the field equations. Thus, we define the decoupled dynamical Gauss-Bonnet (D$^2$GB) gravity~\cite{Yagi:2015oca}, where $S_\EDGB$ is replaced with
\be
S_\DDGB = \alpha_\GB \,  \int d^4x \sqrt{-g} \; \phi \; R_\GB^2\,.
\ee
This theory is shift-invariant, in the sense that the field equations
are invariant under a constant shift in the scalar field. In fact,
this theory is a special case of Horndeski
gravity~\cite{Kobayashi:2011nu}, the most generic class of
scalar-tensor theories with second-order field
equations~\cite{Horndeski:1974wa,Deffayet:2009mn,Deffayet:2011gz} as
already explained in Sec.~\ref{subsec:horndeski}. More specifically it
falls within the class of shift-symmetric Horndeski theories, and it
is free of Ostrogradski ghosts~\cite{Crisostomi:2017ugk}.  The
well-posedness of the theory is still an open question:
Papallo~\cite{Papallo:2017ddx} proved that D$^2$GB gravity fails to be
strongly hyperbolic in any generalized harmonic gauge.  Black hole
solutions in D$^2$GB gravity have been constructed \emph{analytically}
for non-rotating
configurations~\cite{Mignemi:1992nt,Mignemi:1993ce,Yunes:2011we,Sotiriou:2014pfa}
and slowly rotating
ones~\cite{Pani:2011gy,Ayzenberg:2014aka,Maselli:2015tta}.

Deriving (monopole) scalar charges, also known as
``sensitivities''~\cite{Will:1989sk}, is crucial to understanding the
emission of scalar radiation. One can extract such scalar charges by
studying the asymptotic behavior of the scalar field at spatial
infinity. So far, scalar charges have only been discussed in D$^2$GB
gravity (but see~\cite{Yagi:2015oca} for approximate calculations of
stellar scalar charges in EdGB gravity). Scalar charges for black
holes were extracted for nonrotating
configurations~\cite{Yagi:2011xp,Sotiriou:2014pfa} and one can easily
extract charges for slowly rotating black holes: see
e.g.~\cite{Ayzenberg:2014aka}. In Appendix~\ref{app:scalar-charge}, we
derive the scalar charge for black holes with arbitrary rotation. The
final expression was shown in~\cite{Yunes:2016jcc}, but the details of
the derivation (together with the quadrupole scalar charge expression)
are presented here for the first time. On the other hand, scalar
charges for ordinary stars
vanish~\cite{Yagi:2011xp,Yagi:2015oca}. This can be shown by
integrating the scalar field equation
\be
\label{eq:D2GB-scalar-eq}
\Box \phi = - \alpha_\GB\, R_\GB^2
\ee
over the entire spacetime. The left-hand side gives the scalar charge, while the right-hand side generates a topological term and boundary terms after applying the generalized Gauss-Bonnet-Chern theorem~\cite{Alty:1994xj,Gilkey:2014wca}. Both of these vanish for simply connected, stationary spacetimes (i.e., for stars). This means that when a star collapses to a black hole, it suddenly \emph{acquires} a scalar charge in a nontrivial manner: this is the opposite of what happens in scalar-tensor theories, where stars \emph{lose} their scalar charge during collapse to a black hole. Such a nontrivial process was confirmed numerically under an Oppenheimer-Snyder collapse for nonrotating configurations~\cite{Benkel:2016rlz} (see also~\cite{Benkel:2016kcq} for related work). One can in fact show that scalar charges for stars vanish more in general for shift-symmetric Horndeski theories~\cite{Barausse:2015wia,Barausse:2017gip,Lehebel:2017fag}, which include D$^2$GB gravity as a special class. This argument fails in the absence of shift symmetry~\cite{Silva:2017uqg,Doneva:2017duq,Antoniou:2017acq,Antoniou:2017hxj}. 

%------------------------------------------------------------------
\subsection{Dynamical Chern-Simons Gravity}
%[Nico]

Dynamical Chern-Simons (dCS) gravity is an effective field theory model that modifies the Einstein-Hilbert action through a quadratic-curvature correction~\cite{Alexander:2009tp}. Motivated from the Green-Schwarz anomaly-canceling mechanism in field theory and its higher-dimensional generalization in string theory~\cite{Alexander:2004xd}, as well as from loop quantum gravity upon scalarization of the Barbero-Immirzi parameter~\cite{Taveras:2008yf,Calcagni:2009xz} and from effective field theories of inflation~\cite{Weinberg:2008hq}, the dCS action is 
\be
S = S_\EH + S_\dCS + S_\phi + S_\mrm{mat}\,,
\ee
where $S_\mrm{mat}$ is the matter action, while $S_\dCS$ and $S_\phi$ are a quadratic-curvature interaction term and the kinetic term for the scalar field $\phi$ respectively, namely
\be
S_\dCS  = \alpha_\dCS \,  \int d^4x \sqrt{-g} \; f(\phi) \; ^{*}RR\,, \quad
S_\phi = - \frac{1}{2} \int d^4 x \sqrt{-g} \; \nabla_\mu \phi \; \nabla^\mu \phi\,.
\ee
Here $\alpha_\dCS$ is a coupling constant and
\be
^{*}RR \equiv {}^{*}R_{\mu\nu\rho\sigma}R^{\nu\mu\rho\sigma}
\ee
is the Pontryagin density, with 
\be
^{*}R_{\mu\nu\rho\sigma} \equiv \frac{1}{2} \epsilon_{\rho\sigma}{}^{\alpha \beta} R_{\mu\nu\alpha \beta}
\ee
the dual Riemann tensor. We assume here that the scalar field is dimensionless in geometrical units, which forces $\alpha_\dCS$ to have units of length squared. For simplicity, we assume we can Taylor expand $f(\phi) = \phi$ to leading non-vanishing order. The constant term in the Taylor expansion does not contribute, because the Pontryagin density is a topological invariant, i.e.~its integral over the manifold is the Pontryagin number, which is related to the manifold's winding number. This means that the Pontryagin density can be written as the four-divergence of a four-current, which becomes a boundary term upon integration by parts and does not contribute to the field equations. The resulting theory with $f(\phi) = \phi$ is then naturally shift-invariant, and in order to preserve this symmetry one typically sets the scalar field potential to zero. 

The dCS field equations can be obtained by varying the action with respect to the metric and the scalar field. The resulting field equations, however, must be understood as effective, given that they arise from an effective action. In particular, this means that $S_{\dCS}$ must be much smaller than $S_{\EH}$ during any process considered; if this were not the case, then higher-order curvature terms would have to be included in the action, which would then modify the field equations upon variation. As an effective field theory, it is trivial to prove that dCS gravity is well-posed~\cite{Delsate:2014hba}. If one insists in treating the theory as exact, however, the appearance of third-order time derivatives in the field equations probably renders this exact version ill-posed~\cite{Delsate:2014hba}. Spinning black hole solutions in dCS gravity have been found \emph{analytically} for slowly rotating black holes~\cite{Yunes:2009hc,Konno:2009kg,Yagi:2012ya} and for black holes in the extremal limit~\cite{Stein:2014xba,McNees:2015srl}, both sets of which differ from the Kerr metric family. Non-spinning black holes, and in fact, any static and spherically symmetric matter configuration have zero dCS modification because the Pontryagin density vanishes identically~\cite{Jackiw:2003pm,Yunes:2007ss,Grumiller:2007rv}. Current constraints on $\alpha_\dCS$ come from Solar System and table top experiments, but they are extremely weak due to the weak fields in the Solar System~\cite{alihaimoud-chen,Yagi:2012ya}. 

Just as in the case of EdGB gravity, the calculation of (monopole) scalar charges (or sensitivities) is also crucial here to understand the emission of scalar radiation. The asymptotic behavior of the scalar field at spatial infinity has revealed that black holes have a rather large scalar charge~\cite{Yunes:2009hc,Yagi:2012ya}, while the scalar charge of neutron stars is greatly suppressed~\cite{Yagi:2011xp,Yagi:2013mbt}; the derivation of the latter result follows closely the explanation in EdGB gravity presented in the previous section. As in the EdGB case, this means that stars have a tiny scalar field, but this grows upon gravitational collapse, until the charge asymptotes that of isolated black holes. 

%------------------------------------------------------------------
\subsection{Lorentz-violating Gravity}
%[Kent]

Lorentz symmetry is one of the fundamental building blocks of GR. Some modified theories of gravity break this symmetry in order to provide a power-counting renormalizable completion of GR in the ultra-violet regime, as proposed by Ho\v rava~\cite{Horava:2009uw} (see~\cite{Wang:2017brl} for a recent review). Lorentz symmetry breaking in the matter sector has been constrained very stringently from, e.g., particle physics experiments~\cite{Kostelecky:2003fs,Kostelecky:2008ts,Mattingly:2005re,Jacobson:2005bg}. A model-independent framework called the Standard Model Extension (SME)~\cite{Colladay:1998fq,Kostelecky:1998id,Kostelecky:1999rh} was developed to map various observations to bounds on Lorentz symmetry breaking. This framework is efficient to probe such a breakage in the matter sector~\cite{Kostelecky:2008ts} or in the sector where matter directly couples with gravity~\cite{Kostelecky:2010ze}. On the other hand, Lorentz symmetry breaking in the gravity sector has not been well constrained, and different mechanisms exist to prevent the breakage in the matter sector to percolate into the gravity sector (see e.g.~\cite{Liberati:2013xla,Pospelov:2010mp}). Bounds on gravitational Lorentz symmetry within the SME context can be found in~\cite{Bailey:2006fd,Bailey:2009me,Bailey:2013oda}.

One example of Lorentz-violating gravity is Einstein-\AE ther (EA) theory~\cite{Jacobson:2000xp,Jacobson:2008aj}, which breaks the gravitational Lorentz symmetry by introducing a preferred time direction at each point in spacetime via a timelike unit vector $U^\mu$. One can consider this theory as a low-energy effective theory of some unknown dynamics at high energy~\cite{Blas:2009qj}. This theory is the most general vector-tensor model whose action only depends on a unit timelike vector field (\AE ther field) and its first derivative, and is quadratic in the latter. Such an action is given by $S = S_\EH + S_{\AE}$, where
\be
S_{\AE} = - \frac{\kappa}{G_{\AE}} \int d^4x \sqrt{-g} M^{\alpha\beta}_{\mu\nu} \nabla_\alpha U^\mu \nabla_\beta U^\nu\,
\ee
and
\be
M^{\alpha\beta}_{\mu\nu} = c_1 g^{\alpha\beta} g_{\mu\nu} + c_2 \delta^\alpha_\mu \delta^\beta_\nu + c_3 \delta^\alpha_\nu \delta^\beta_\mu +c_4 U^\alpha U^\beta g_{\mu\nu}\,.
\ee
Here $\kappa \equiv 1/16\pi$, $G_{\AE}$ is the bare gravitational constant in the theory, while $c_i$ $(i=1...4)$ are coupling constants. Solar System experiments essentially reduce the parameter space from four to two, $c_{\pm} = c_1 \pm c_3$~\cite{Jacobson:2008aj}. These remaining parameters have been constrained from binary pulsar observations~\cite{Yagi:2013qpa,Yagi:2013ava} and from the recent coincident GW and electromagnetic observation of merging neutron stars~\cite{Monitor:2017mdv}. The binary pulsar bounds are derived by calculating the sensitivities of neutron stars, which are obtained by constructing a slowly moving neutron star solutions with respect to the \AE ther field (black hole sensitivities have not been calculated yet). The gravitational wave bounds come from the speed of the propagating degrees of freedom in EA theory, the two tensor modes, the two vector modes and the one scalar mode~\cite{Jacobson:2004ts,Foster:2005dk,Foster:2006az,Foster:2007gr}. Tensor, vector and scalar modes propagate at speeds $w_2^{\AE}$, $w_1^{\AE}$ and $w_0^{\AE}$ respectively, which can be expressed in terms of the coupling constants, as summarized in Table~\ref{table:EA-KG-prop-speed}. Here, we defined $c_{14} \equiv c_1 + c_4$ and $c_{123} \equiv c_1+c_2+c_3$.

{
\newcommand{\minitab}[2][l]{\begin{tabular}{#1}#2\end{tabular}}
\renewcommand{\arraystretch}{2.}
%\begingroup
%\squeezetable
\begin{table*}[htb]
\begin{centering}
\begin{tabular}{c|c|c}
\hline
\hline
\noalign{\smallskip}
Theory & mode & propagation speed \\
\hline  \hline
\multirow{3}{*}{Einstein-\AE ther} & tensor & $w_2^{\AE} = \frac{1}{1-c_+} $  \\
& vector & $w_1^{\AE} = \frac{2c_1-c_+c_-}{2(1-c_+)c_{14}}$  \\
& scalar & $w_0^{\AE} = \frac{(2-c_{14})c_{123}}{(2+3c_2+c_+)(1-c_+)c_{14}}$  \\
\hline
\multirow{2}{*}{khronometric} & tensor & $w_2^{\KG} = \frac{1}{1-\beta} $  \\
& scalar & $w_0^{\KG} = \frac{(2-\alpha)(\beta + \lambda)}{(2+3\lambda+\beta)(1-\beta)\alpha}$  \\
\noalign{\smallskip}
\hline
\hline
\end{tabular}
\end{centering}
\caption{
Propagation speed of tensor, vector and scalar modes in EA and khronometric theory. Vector modes are absent in the latter.
 }
\label{table:EA-KG-prop-speed}
\end{table*}
%\endgroup
}

Another example of Lorentz-violating gravity that is related to EA theory is khronometric gravity~\cite{Blas:2010hb}. This theory corresponds to the low-energy limit of Ho\v rava gravity. The action is the same as the EA case, but now the \AE ther field is hypersurface orthogonal. In other words, this theory introduces a preferred time foliation of spacetime (or a global preferred time) and the \AE ther field is normal to the hypersurface of constant preferred time. This reduces the number of coupling constants from four to three: $\lambda_\KG = c_2$, $\beta_\KG = c_1 + c_3$ and $\alpha_\KG = c_1 + c_4$. Imposing the Solar System bounds essentially eliminates $\alpha_\KG$. One can further constrain the remaining parameters from binary pulsar~\cite{Yagi:2013qpa,Yagi:2013ava}, Big Bang nucleosynthesis~\cite{Audren:2013dwa} and gravitational-wave~\cite{Monitor:2017mdv,Gumrukcuoglu:2017ijh} observations. Regarding the propagation degrees of freedom in khronometric gravity, there are two tensor modes and one scalar mode~\cite{Blas:2011zd}, whose speeds are given by $w_2^{\KG}$ and $w_0^{\KG}$, respectively. Their explicit forms are also summarized in Table~\ref{table:EA-KG-prop-speed}.

%------------------------------------------------------------------
\subsection{Extra Dimensional Theories}

String theory predicts that we live in a higher-dimensional spacetime, and extra dimensions are compactified. One simple example of such a compactification is the Kaluza-Klein compactification. Particle physics experiments place bounds on the size $\ell$ of the extra dimension: $\ell < 10^{16}$cm. Arkani-Hamed, Dimopoulos and Dvali (ADD)~\cite{ArkaniHamed:1998rs,ArkaniHamed:1998nn} proposed a braneworld model with a tensionless brane (on which we live) embedded in a flat and compact bulk spacetime. In the ADD model matter is localized on the brane, so that only gravitons can propagate through the bulk. The size of the extra dimensions can be relatively large in this model, since bounds on the gravity sector are much weaker than those on the matter sector. Furthermore, this model can naturally explain the hierarchy problem between the electroweak and Planck scale. 

Another braneworld model was proposed by Randall and Sundrum. The first model (RS-I)~\cite{Randall:1999ee} introduces one positive-tension and one negative-tension brane in a five-dimensional anti-de Sitter (AdS) bulk spacetime. With this model, one can choose the separation of two branes such that the fundamental five-dimensional Planck mass is $\sim \ell^{-1} \sim 1$TeV, while the four-dimensional Planck mass becomes $10^{19}$GeV. In their second model (RS-II)~\cite{Randall:1999vf}, they take the negative-tension brane to infinity, thus the model effectively includes only one brane. The remarkable feature of this model is that Newtonian gravity is reproduced in the low-energy limit~\cite{Garriga:1999yh} even if the size of the extra dimension is relatively large. Table-top experiments place the bound $\ell < 14\mu$m~\cite{Adelberger:2006dh}. 
 
Since the bulk of the RS-II model is AdS, one can apply the AdS/CFT conjecture~\cite{Maldacena:1997re,Aharony:1999ti}, which states that the four-dimensional $\mathcal{N}=4$ $U(N)$ super Yang-Mills theory on the AdS boundary can be recovered from gravity in the $AdS_5 \times S^5$ spacetime. In particular, one can apply this conjecture to brane-localized black holes. In the CFT picture, due to the large number of CFT degrees of freedom~\cite{Aharony:1999ti}, four-dimensional black holes on the brane evaporate significantly faster than in GR. In the AdS picture, this is seen as \emph{classical} evaporation of a black hole of mass $M$~\cite{emparan-conj,tanaka-conj}, with evaporation rate given by~\cite{Emparan:2002jp}
\be
\label{eq:dmdt}
\frac{dM}{dt} = -2.8 \times 10^{-7} \left( \frac{1M_\odot}{M} \right)^2 \left( \frac{\ell}{10\mu \mrm{m}} \right)^2 \frac{M_\odot}{\mrm{yr}}\,.
\ee
Later, static brane-localized black hole solutions were found numerically~\cite{Figueras:2011gd,Abdolrahimi:2012qi}, which questions the validity of this conjecture. Nevertheless, we consider placing bounds on $\ell$ assuming that the conjecture is correct. This is because one can easily map this mass loss effect of black holes in the RS-II model to a similar effect due to (e.g.) phantom energy accretion in GR~\cite{Babichev:2004yx,Babichev:2005py,Babichev:2014lda}. In terms of GWs, a change in the black hole mass is similar to a change in the gravitational constant $G$~\cite{Yunes:2009bv}, which is predicted in many modified theories of gravity.

%%%%%%%%%%%%%%%%%%%%%%%%%%%%%%%
\section{Inspiral Tests of Modified Gravity}
\label{sec:inspiral}

%------------------------------------------------------------------
\subsection{Generic Tests}

The plethora of modified gravity models that have been proposed and the extreme difficulty in constructing sufficiently accurate GW models for data analysis suggests that theories ought not to be treated on a case-by-case basis. Indeed, it took the gravity theory community approximately 50 years to obtain a sufficiently accurate (third post-Newtonian (PN) order) model of the GWs emitted in the inspiral of compact binaries in GR. Instead of carrying out similar calculations on a theory-by-theory basis, it is much more appealing to develop generic tests of Einstein's theory given the available data. Indeed, a similar approach was successfully pursued when carrying out tests with Solar System observations, which led to the development of the parameterized PN framework of Will and Nordtvedt~\cite{Nordtvedt:1968qs,1971ApJ...163..611W,1972ApJ...177..757W,1972ApJ...177..775N}.  

The first attempt at such a generic test consisted of verifying the PN structure of the waveform phase~\cite{Arun:2006yw}. The idea was to decompose the Fourier-domain waveform model into a frequency-dependent amplitude and a frequency-dependent phase, and to then rewrite the phase as\footnote{The terms $\alpha_5$ and $\alpha_6$ contain contributions that depend on $\ln v$, which the authors treat as constant in~\cite{Arun:2006yw}. In their follow-up papers~\cite{Arun:2006hn,mishra}, they modified Eq.~\eqref{eq:phase-arun} by adding further terms of the form $\sum_{k} \alpha_{n,l} \ln v$.}
\be
\label{eq:phase-arun}
\Psi(f) = \sum_{n=0}^{n=7} \alpha_{n} v(f)^{-5+n}\,,
\ee
where $\alpha_{n}$ are PN coefficients, which in GR are known functions of the parameters of the binary (to be more precise , the individual masses $m_1$ and $m_2$ for non-spinning black hole binaries), and $v(f) = (\pi m f)^{1/3}$ is the orbital velocity, with $m$ the binary's total mass. The proposal was then to treat all of these coefficients as independent and find the best-fit values by comparing the above template waveform with the data. One can then draw error regions of each coefficient in the $m_1$-$m_2$ plane assuming GR is correct to check for consistency, namely to check if there is a region where \emph{all} of error regions overlap. Later the authors only considered three out of eight coefficients, so that correlations among parameters could be reduced and one could carry out a stronger test by shrinking the error regions~\cite{Arun:2006hn,mishra}. This procedure resembles binary pulsar tests in the parameterized post-Keplerian formalism~\cite{Damour:1991rd,Stairs:2003eg}.

Although feasible in principle, the above test has a few limitations. First, it has the strong bias of assuming Nature follows the same exact functional structure of the PN approximation in GR, i.e.~that the Fourier phase can be expressed as a series in integer powers of velocity, with the leading-order term starting at $v^{-5}$. Indeed, many examples of modified gravity effects and modified gravity theories exist which do not admit this structure; examples of this include dipole emission ($\propto v^{-7}$), variability of the fundamental constants ($\propto v^{-13}$), parity violation in eccentric binaries ($\propto v^{-7.3}$), and massive gravitons in eccentric binaries ($\propto v^{-9.3}$), to name a few. Second, the framework does not allow for tests of modified gravity theories that lead predominantly to amplitude modifications, without affecting the phase evolution much; examples of this include gravitational birefringence~\cite{Alexander:2007kv,Yunes:2010yf,Yagi:2017zhb}. Third, the framework assumes that polynomials in velocity are a good basis to expand the Fourier phase during the entire inspiral, including right up to plunge and merger. Today, we know that this is not the case, with the series requiring arctangent  corrections~\cite{Husa:2015iqa,Khan:2015jqa}. 

An extension and generalization of this method that resolves all of the above problems is the parameterized post-Einsteinian (ppE) approach~\cite{Yunes:2009ke}. In this framework, one extends the GR waveform model via
\be
\tilde{h}(f) = \tilde{A}_{\GR}(f) \left[1 + \alpha_{\ppE}\, v(f)^{a}\right] e^{i \Psi_{\GR}(f) + i \beta_{\ppE}\, v(f)^{b}}\,,
\ee
where $\tilde{A}_{\GR}(f)$ and $\Psi_{\GR}(f)$ are the Fourier amplitude and Fourier phase in the most accurate GR model developed at that stage in time. The quantities $(\alpha_{\ppE},\beta_{\ppE})$ are ppE constants that control the magnitude of deviations from GR, while $(a,b)$ are real numbers that determine the type of deviation that is being constrained. This parameterization resolves the issues listed above because (i) it allows for deformations from GR that enter at pre-Newtonian order, i.e.~in negative powers of velocity relative to the leading-order PN term in GR, (ii) it allows for both amplitude and phase deformations, and (iii) it allows the deformation of the most sophisticated waveform model in GR, such that when the ppE parameters are zero, one recovers exactly the most accurate model. Different variants of this approximation have been and can be used, with for example different scalings in the ppE terms, working entirely in the time domain, or working with multiple amplitude and phase ppE parameters at the same time~\cite{Cornish:2011ys,Chatziioannou:2012rf,Sampson:2013jpa,Sampson:2013lpa}.

The idea of the ppE framework is then to use the above model and let the data decide (the posteriors for) the magnitude of the ppE parameters $(\alpha_{\ppE},\beta_{\ppE})$ for whatever choice of $(a,b)$ modification one wishes to study. Once these posteriors have been constructed, one can then \emph{map} them to posteriors on coupling parameters of specific theories. This mapping step is crucial because it is what allows us to draw inferences on different specific aspects of theoretical physics from the observations, as done recently with the first GW observations~\cite{Yunes:2016jcc}. This step, of course, requires that one first calculate predictions for the GWs emitted by compact binaries in specific theories. These predictions have been obtained in many theories, as we detail in the next subsection. The ppE parameterization also relaxes the \emph{a priori} assumption that GR is correct, allowing the data itself to select the theory that best supports it, thus minimizing what has been dubbed as \emph{theoretical bias}~\cite{Yunes:2009ke,Vallisneri:2013rc}. Fortunately, all GW observations to date seem consistent with GR, implying that theoretical bias is not a problem right now. 

Inferences drawn from constraints on ppE parameters can be classified into two classes, depending on the sector of the theory they constrain: \emph{generation} and \emph{propagation}. The generation sector deals with the way the particular theory sources GWs, and any additional degree of freedom, and how these evolve in time and back-react on the evolution of the binary system. The propagation sector deals with how the waves travel away from the binary system to Earth, according to the wave's dispersion relation. One can, of course, devise more generic parameterizations that account for more exotic phenomena during the generation and propagation of GWs. For example, one could imagine that an additional massive degree of freedom is present in the theory, leading to modifications to the dynamics only once the binary radiates at the Compton wavelength  of the new degree of freedom. This would lead to a sharp turn-on and turn-off of GR modifications that are not captured by the parameterization above. However, data analysis investigations have revealed that even if Nature were that cruel, the ppE parameterization presented above would still be able to detect an anomaly in the signal, although in a suboptimal way~\cite{Sampson:2013lpa}. Similarly, one could imagine that in a yet-to-be-discovered modified theory GWs propagate differently depending on the direction in which they travel~\cite{Tso:2016mvv}. Such a direction-dependent dispersion relation would not fit perfectly within the ppE parameterization, but a signal of that type can be recovered with the simplest ppE model written above, although sub-optimally.    

A recent idea has been put forth to draw generation-type inferences on the nature of the black holes that generate GWs in a binary. The idea is to allow for the quadrupole moment of each black hole in a binary to be a new parameter in the waveform model, and then to check whether the measured parameter agrees with the GR expectation per the no-hair theorems~\cite{Krishnendu:2017shb}. A generic quadrupole in the waveform will introduce a second PN order correction to the GW phase, which thus maps in a one-to-one fashion to the ppE parameterization described above. This correction, however, will be degenerate with the spins of the black holes, so unless these can be separately extracted, it would be difficult to implement this test. The independent extraction of the spins can be achieved if the signal and the waveform model include spin-orbit and spin-spin precession effects, which induce amplitude modulations that can in principle break this degeneracy~\cite{Chatziioannou:2014coa,Chatziioannou:2014bma,Chatziioannou:2016ezg,Chatziioannou:2017tdw}.

More in general, the binary dynamics of hypothetical black hole alternatives in binary systems is driven by their so-called tidal Love numbers. Tidal Love numbers encode the deformability of a self-gravitating object immersed in a tidal environment, and they depend both on the object's internal structure and on the dynamics of the gravitational field. In classical general relativity, the tidal Love numbers of black holes are exactly zero. Recent work computed the tidal Love numbers of various exotic compact objects, including boson stars, gravastars and wormholes, as well as black holes in various theories of modified gravity (Einstein-Maxwell, Brans-Dicke and Chern-Simons gravity)~\cite{Cardoso:2017cfl}. In general, these calculations showed that the tidal Love numbers of exotic objects depend logarithmically on the location of the putative ``surface'' replacing the black hole's horizon. LIGO-like detectors, third-generation Earth-based detectors and LISA can impose interesting constraints on the tidal Love numbers of boson stars, while a LISA-like detector could probe even extremely compact objects~\cite{Maselli:2017cmm}, as long as systematic errors in general relativity are under control.

{
\newcommand{\minitab}[2][l]{\begin{tabular}{#1}#2\end{tabular}}
\renewcommand{\arraystretch}{2.}
%\begingroup
%\squeezetable
\begin{table*}[thb]
\begin{centering}
\begin{tabular}{c|c|c}
\hline
\hline
\noalign{\smallskip}
Theory & $\beta_\ppE$ & $b$ \\
\hline  \hline
scalar-tensor~\cite{Jacobson:1999vr,Horbatsch:2011ye,Yunes:2016jcc} & $- \frac{5}{1792} \dot{\phi}^2 \eta^{2/5} \left(m_1 s_1^\ST - m_2 s_2^\ST \right)^2$ & $-7$ \\
EdGB, D$^2$GB~\cite{Yagi:2011xp} & $- \frac{5}{7168} \zeta_{\GB} \frac{\left(m_1^2 s_2^\GB - m_2^2 s_1^\GB \right)^2}{m^4 \eta^{18/5}}$ & $-7$ \\
dCS~\cite{Yagi:2012vf} & $\frac{1549225}{11812864} \frac{\zeta_{\CS}}{\eta^{14/5}} \left[\left(1 - \frac{231808}{61969} \eta\right) \chi_{s}^{2} 
+ \left(1 - \frac{16068}{61969} \eta\right) \chi_{a}^{2} - 2 \delta_{m} \chi_{s} \chi_{a}\right]$ & $-1$ \\
EA~\cite{Hansen:2014ewa} & $- \frac{3}{128} \left[ \left( 1-\frac{c_{14}}{2} \right) \left(\frac{1}{w_2^{\AE}}  +\frac{2c_{14}c_+^2}{(c_++c_--c_-c_+)^2 w_1^{\AE}}
	+\frac{3c_{14}}{2w_0^{\AE} (2-c_{14})} \right) - 1 \right]$ & $-5$ \\
khronometric~\cite{Hansen:2014ewa} & $- \frac{3}{128} \left[ \left( 1 - \beta_\KG \right) \left(\frac{1}{w_2^{\KG}} 
	\frac{3 \beta_\KG}{2 w_0^{\KG} (1- \beta_\KG)} \right) - 1 \right]$ & $-5$ \\
extra dimension~\cite{yagi:brane} & $ \frac{25}{851968} \left(\frac{dm}{dt}\right)  \frac{3 - 26 \eta + 34 \eta^{2}}{\eta^{2/5} (1-2\eta)}$ & $-13$ \\
varying $G$~\cite{Yunes:2009bv} & $- \frac{25}{65536} \dot{G} {\cal{M}}$ & $-13$ \\
mod. disp. rel.~\cite{Mirshekari:2011yq} & $\frac{\pi^{2-\alpha_\MD}}{(1-\alpha_\MD)} \frac{D_{\alpha_\MD}}{\lambda_{\mathbb{A}}^{2-\alpha_\MD}}  \frac{\mathcal{M}^{1-\alpha_\MD}}{(1+z)^{1-\alpha_\MD}}$ & $3 (\alpha_\MD-1)$  \\
\noalign{\smallskip}
\hline
\hline
\end{tabular}
\end{centering}
\caption{
Mapping of ppE parameters to those in each theory for a black hole binary. In scalar-tensor theories, black holes acquires a scalar charge for a cosmologically evolving scalar field~\cite{Jacobson:1999vr,Horbatsch:2011ye}. Such a scalar charge is proportional to $s_A^\ST \equiv  [1+(1-\chi_A^2)^{1/2}]/2$. $s_A^\GB$ is related to the black hole scalar charge $\mu$ in D$^2$GB in Eq.~\eqref{eq:scalar-charge-D2GB} as $\mu_A^\GB = 2 (\alpha_\GB/m_A^2) s_A^\GB$. The dimensionless coupling constant in quadratic-curvature theories is defined by $\zeta_{\GB,\CS} = 16 \pi \alpha_{\GB,\CS}^2/m^4$. Propagation speeds $w_i^{\AE,\KG}$ in Lorentz-violating theories are summarized in Table~\ref{table:EA-KG-prop-speed}. $dm/dt = dm_1/dt + dm_2/dt$ can be calculated from Eq.~\eqref{eq:dmdt}. $\lambda_{\mathbb{A}} \equiv h \, \mathbb{A}^{1/(\alpha-2)}$, where $h$ is the Planck constant. The distance $D_{\alpha_\MD}$ is defined in Eq.~\eqref{eq:D_alpha}. 
 }
\label{tab:mapping-ppE}
\end{table*}
%\endgroup
}

%------------------------------------------------------------------
\subsection{Mapping to Specific Theories}

The mapping between the ppE parameters and the coupling constants in various modified theories of gravity is summarized in Table~\ref{tab:mapping-ppE}. Since matched filtering data analysis is more sensitive to deviations in the waveform phase than the amplitude, we only show the mapping for the phase. The mapping in the table corresponds to GWs from black hole binaries with component masses $m_A$ and dimensionless spins $\chi_A$ ($A=1,2$). We also introduce the total mass $m=m_1+m_2$, symmetric mass ratio $\eta = m_1 m_2/m^2$ and chirp mass $\mathcal M = m \eta^{3/5}$, as well as two (symmetric and antisymmetric) spin parameters $\chi_s = (\chi_1 + \chi_2)/2$ and $\chi_a = (\chi_1 - \chi_2)/2$.

Of all the theories considered in the inspiral of compact objects, scalar-tensor theories are by far the most well-studied. Pioneering studies of compact binaries in scalar-tensor theory were
carried out by Eardley~\cite{1975ApJ...196L..59E}, Will and Zaglauer~\cite{Will:1977wq,Will:1989sk} and Damour and Esposito-Far\`{e}se~\cite{Damour:1992we,Damour:1996ke,Damour:1998jk}, among others. More recently, the post-Minkowskian technique of direct integration of the relaxed Einstein equations (DIRE)~\cite{Will:1996zj,Pati:2000vt,Pati:2002ux} was used to compute the equations of motion for a system of compact objects at 2.5PN order, as well as the gravitational waveform and energy flux at 2PN (relative) order for binaries on generic orbits~\cite{Mirshekari:2013vb,Lang:2013fna,Lang:2014osa}. Sennett et al.~\cite{Sennett:2016klh} specialized the results of~\cite{Mirshekari:2013vb,Lang:2013fna,Lang:2014osa} to non-spinning binary systems on quasi-circular orbits in scalar-tensor gravity at 2PN relative order\footnote{By imposing the stringent constraints set by current astrophysical observations (cf. Table II of~\cite{Sennett:2016klh}), they find that dipolar radiation is subdominant to quadrupolar radiation for most prospective GW sources: {\em in the absence of spontaneous scalarization}, the dipole term can dominate only at frequencies $f\lesssim 100~\mu$Hz in binary neutron star or neutron-star/stellar-mass-black-hole systems, and at frequencies $f \lesssim 5~\mu$Hz in neutron-star/intermediate-mass-black-hole systems.  Therefore, ground- and space-based GW detectors would only observe binary systems whose inspiral is driven by the next-to-leading order flux.}. Such waveforms introduce PN corrections to the mapping presented in Table~\ref{tab:mapping-ppE}. Juli\'e and Deruelle~\cite{Julie:2017pkb,Julie:2017ucp} use these higher order PN results to begin to extend the effective-one-body (EOB) formalism of Buonanno and Damour~\cite{Buonanno:1998gg} to scalar-tensor gravity. Such resummed waveform models cannot be analytically mapped to the ppE waveforms directly.   

{
\newcommand{\minitab}[2][l]{\begin{tabular}{#1}#2\end{tabular}}
\renewcommand{\arraystretch}{1.4}
%\begingroup
%\squeezetable
\begin{table*}[htb]
\begin{centering}
\begin{tabular}{c|c|c}
\hline
\hline
\noalign{\smallskip}
Theory & $\mathbb{A}$ & $\alpha_\MD$ \\
\hline  \hline
massive gravity~\cite{Will:1997bb,Rubakov:2008nh,Hinterbichler:2011tt,deRham:2014zqa} & $m_g^2$ & 0 \\
\hline
\multirow{2}{*}{multifractional spacetime~\cite{Calcagni:2009kc,Calcagni:2011kn,Calcagni:2011sz,Calcagni:2016zqv}} & $\frac{2}{3-\alpha_\MD} E_*^{2-\alpha_\MD}$ (timelike spacetime) & \multirow{2}{*}{2--3} \\
& $-\frac{2 \cdot 3^{1-\alpha_\MD/2}}{3-\alpha_\MD} E_*^{2-\alpha_\MD}$ (spacelike spacetime) & \\
\hline
double Special Relativity~\cite{AmelinoCamelia:2000ge,Magueijo:2001cr,AmelinoCamelia:2002wr,AmelinoCamelia:2010pd} & $\eta_\mrm{dsrt}$ & 3 \\
\hline
extra dimension~\cite{Sefiedgar:2010we} & $- \alpha_\mrm{edt}$ & 4 \\
\hline
\multirow{2}{*}{SME~\cite{Kostelecky:2016kfm}} & $- 2 \mathring{k}_{(I)}^{(d)}$ (even $d \geq 4$) & \multirow{2}{*}{$d-2$} \\
& $\pm 2 \mathring{k}_{(V)}^{(d)}$  (odd $d \geq 5$) & \\
\hline
Ho\v rava~\cite{Horava:2008ih,Horava:2009uw,Vacaru:2010rd,Blas:2011zd} & $\kappa_\mrm{hl}^4 \mu_\mrm{hl}^2/16$ & 4 \\
\noalign{\smallskip}
\hline
\hline
\end{tabular}
\end{centering}
\caption{
Mapping between modified dispersion relation parameters for the graviton in Eq.~\eqref{eq:disp-rel} and the parameters of each theory. The meaning of the parameters is as follows. $m_g$: the graviton mass; $E_*$: the characteristic length scale above which spacetime is discrete; $\eta_\mrm{dsrt}$: the characteristic observer-independent length scale; $\alpha_\mrm{edt}$: the square of the Planck length in extra dimensional theories; $\mathring{k}_{(I)}^{(d)}$ and $ \mathring{k}_{(V)}^{(d)}$: parameters controlling the Lorentz-violation operators in SME in the rotation-invariant limit; $\kappa_\mrm{hl}$: a parameter related to the bare gravitational constant; $\mu_\mrm{hl}$: a parameter related to the deformation in the ``detailed balance'' conditions in Ho\v rava gravity.
 }
\label{tab:mod-disp-rel}
\end{table*}
%\endgroup
}

All the mappings in Table~\ref{tab:mapping-ppE} (except for the last one) originate from non-GR effects created at the level of generation of GWs, while such waves in general acquire modifications also at the level of their propagation. The dispersion relation of the graviton in non-GR theories can be expressed in terms of two parameters $\mathbb{A}$ and $\alpha_\MD$ as~\cite{Mirshekari:2011yq} 
\be
\label{eq:disp-rel}
E^2 =  \left(p c\right)^2 + \mathbb{A} \left(p c\right)^{\alpha_\MD}\,.
\ee
We present the mapping between these two parameters and coupling constants in each modified theory of gravity in Table~\ref{tab:mod-disp-rel}. From the above dispersion relation, one finds the group velocity of the graviton
\be
\label{eq:prop-speed}
\frac{v_{g}}{c}  =  \frac{1}{c} \frac{d\omega}{dk}= 1 + \frac{(\alpha_\MD -1)}{2} \mathbb{A} \, E^{\alpha_\MD-2}\,.
\ee
The ppE parameters $\beta_\ppE$ and $b$ with the modified dispersion relation of the graviton in Eq.~\eqref{eq:disp-rel} are given by the last line in Table~\ref{tab:mapping-ppE}. Here, the distance parameter $D_{\alpha_\MD}$ is defined by~\cite{Mirshekari:2011yq}
\be
\label{eq:D_alpha}
D_{\alpha_\MD} = \frac{z}{H_{0} \sqrt{\Omega_{M} + \Omega_{\Lambda}}} \left[1 - \frac{z}{4} \left(\frac{3 \Omega_{M}}{\Omega_{M} + \Omega_{\Lambda}} + 2 \alpha_\MD \right) + {\cal{O}}(z^{2})\right]\,,
\ee
where $H_0$ is the local Hubble parameter, $z$ represents the redshift, and $\Omega_{M}$ and $\Omega_{\Lambda}$ are the energy density of dark matter and dark energy, respectively.

%------------------------------------------------------------------
\subsection{Current and Future Tests}

\subsubsection{Current Tests}

The LIGO/Virgo Collaboration (LVC) used the observed data to perform various model-independent tests of gravity in the extreme field regime~\cite{TheLIGOScientific:2016src}. The first test that they performed was to estimate the residual signal-to-noise ratio between the GW150914 data and the GR waveform template. They concluded that GR violations in GW150914 that cannot be absorbed into a redefinition of binary parameters have been constrained to less than 4\%. The second test was to check the consistency in the measurement of the final black hole mass and spin using the inspiral and post-inspiral data. The third test was to look for non-tensorial polarization modes. The presence of such polarization modes was inconclusive for the first three events due to the near-alignment of the two aLIGO detectors. Since Virgo also detected signals for the fourth event GW170814~\cite{Abbott:2017oio}, the LVC could carry out a meaningful test of GR with GW polarizations for the first time. They concluded that the data favors purely tensor polarizations over purely scalar (vector) polarizations with a Bayes factor larger than 1000 (200), respectively. The fourth test was to constrain deviations in the waveform phase away from GR at each PN order. 

\begin{figure}[h]
\begin{center}
\includegraphics[width=8cm,clip=true]{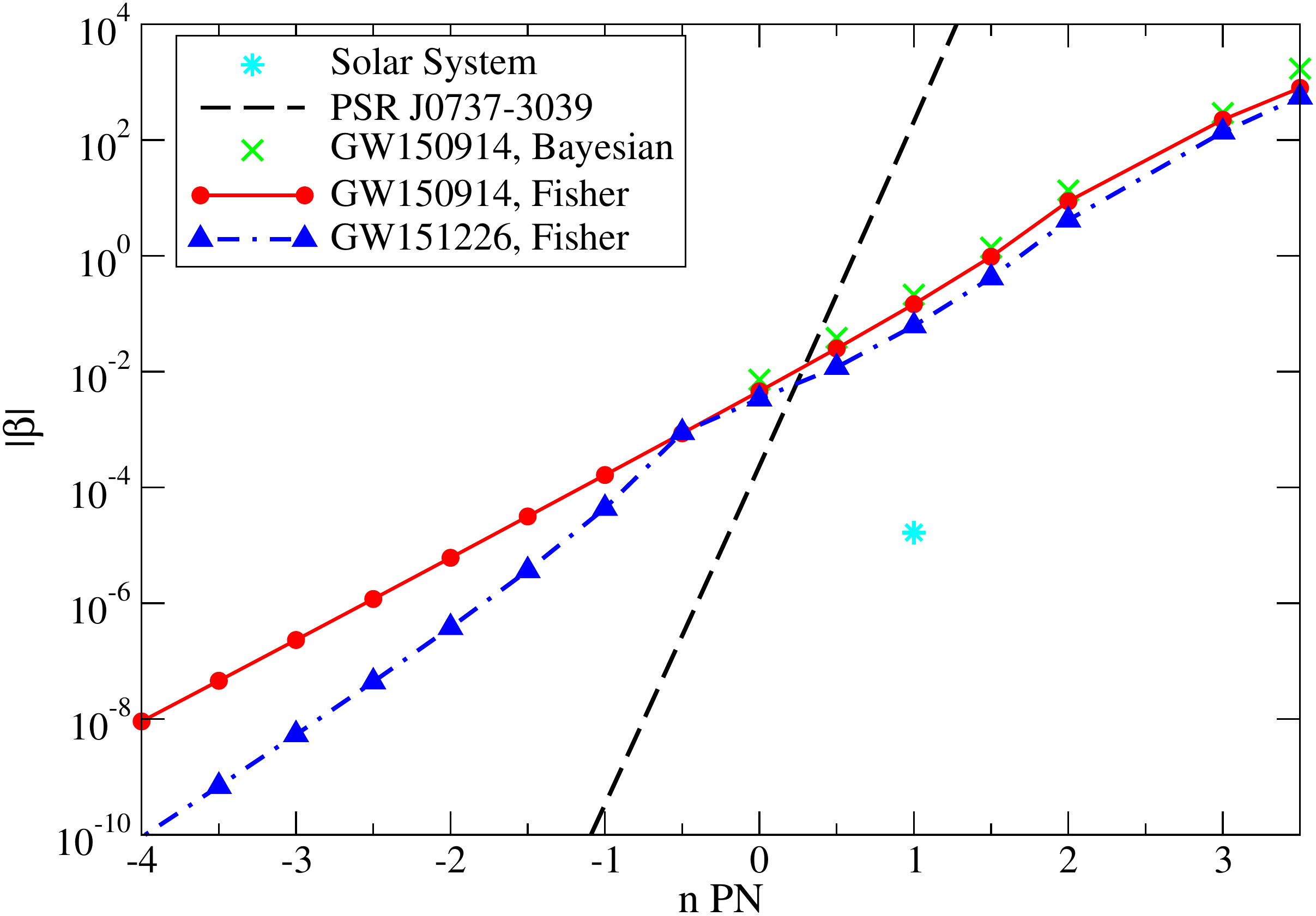} 
\caption{\label{fig:beta-cons} [Adapted from~\cite{Yunes:2016jcc}.] 90\%-confidence constraints on the ppE parameter $|\beta_\ppE|$ entering at $n$th PN order. The green crosses represent the bounds reported in~\cite{TheLIGOScientific:2016pea,TheLIGOScientific:2016src} using a Bayesian analysis of event GW150914. The red (blue) dots and line represent bounds from GW150914 (GW151226) estimated using a Fisher analysis. The dashed black line and the cyan star corresponds to bounds from binary pulsar observations~\cite{Yunes:2010qb} and Solar System experiments~\cite{Sampson:2013wia}. Observe that the GW bounds are stronger than binary pulsar bounds on $\beta_\ppE$ entering at a positive PN order.
}
\end{center}
\end{figure}

The last test mentioned in the previous paragraph can easily be mapped to a test with ppE waveforms. Green crosses in Fig.~\ref{fig:beta-cons} show the GW150914 bound on $\beta_\ppE$ at each PN order that is obtained by mapping the results in~\cite{TheLIGOScientific:2016src} to ppE parameters. These results constrain non-GR corrections arising from GW generation mechanisms, as the non-GR corrections are introduced only in the inspiral part of the waveform. For the merger-ringdown phase, one needs to carry out more merger simulations in non-GR theories to construct valid parameterized waveforms. Observe that bounds on $\beta_\ppE$ with GW150914 are much stronger than those from the orbital decay rate of the double binary pulsar J0737-3039~\cite{Yunes:2010qb} for positive PN corrections. This is because such positive PN corrections become larger in the stronger field regime, precisely where GW observations are most probing. The cyan star indicates the bound on $\beta_\ppE$ at 1PN order obtained from Solar System experiments. Although such a bound is stronger than the GW bound, these two bounds have a different meaning: the former probes weak, non-dynamical gravity, while the latter probes the strong and dynamical regime.

Reference~\cite{Yunes:2016jcc} extended this result by deriving GW bounds also for negative PN corrections\footnote{The LVC derived bounds on the $-1$PN term with GW170814~\cite{Abbott:2017oio}.}. Instead of carrying out a full Bayesian analysis using the actual data, the authors carried out a simpler calculation using a Fisher analysis. They injected a GR waveform whose parameters are consistent with those reported in~\cite{TheLIGOScientific:2016pea}, and recovered them with ppE templates. Bounds on $\beta_\ppE$ for both positive and negative PN corrections with this method for GW150914 are shown by the red curve and circles. Observe first that the bounds on the positive PN corrections agree nicely with those obtained in~\cite{TheLIGOScientific:2016src}. Such an agreement demonstrates the validity of using a Fisher analysis and a GR waveform for injection. Observe also that bounds for negative PN corrections are much weaker than the binary pulsar bounds. A stronger bound on $\beta_\ppE$ does not necessarily mean a stronger bound on a specific modified theory of gravity. This is because $\beta_\ppE$ depends not only on theoretical coupling constants, but also on binary parameters like masses and spins. Moreover, there are theories like D$^2$GB gravity where non-GR corrections from neutron star binaries are highly suppressed, and thus one needs to use observations from binaries with at least one black hole to constrain the theories more efficiently. The blue dashed curve and triangles present bounds from GW151226. The bounds are stronger than GW150914, especially for negative PN corrections. This is because the masses of GW151226 are smaller than those of GW150914, which means that the velocity of binary components are smaller at a fixed frequency, making the corrections larger at negative PN orders.

{
\newcommand{\minitab}[2][l]{\begin{tabular}{#1}#2\end{tabular}}
\renewcommand{\arraystretch}{1.4}
\begingroup
%\squeezetable
\begin{table*}[t]
\begin{centering}
\begin{tabular}{c|c|c|c|c|c}
\hline
\hline
\noalign{\smallskip}
 Theory & GR Pillar & PN & Repr. Parameters & {\bf{GW150914}}  & Other Bounds \\
\hline \hline
 EdGB, D$^2$GB  & \multirow{2}{*}{SEP}  & \multirow{2}{*}{$-1$}  & $\sqrt{|\alpha_\EDGB|}$ [km] & 
 --- &  $10^7$~\cite{Amendola:2007ni}, 2~\cite{Kanti:1995vq,kent-LMXB,Pani:2009wy}\\
 scalar-tensor  &   & & $|\dot{\phi}|$ [1/sec] & 
 --- &   $10^{-6}$~\cite{Horbatsch:2011ye} \\ 
 \hline
 dCS & SEP, PI  & $+2$ &  $\sqrt{|\alpha_\CS|}$ [km] & --- &  $10^8$~\cite{alihaimoud-chen,Yagi:2012ya}\\ 
\hline 
Einstein-\AE ther  & \multirow{2}{*}{SEP, LI} &  \multirow{2}{*}{$0$} & $(c_+,c_-)$ & $\mathbf{(0.9,2.1)}$  & $(0.03,0.003)$~\cite{Yagi:2013qpa,Yagi:2013ava}\\ 
 khronometric & & & $(\beta_\KG,\lambda_\KG)$ & $\mathbf{(0.42,-)}$ & $(0.005,0.1)$~\cite{Yagi:2013qpa,Yagi:2013ava}\\ 
\hline 
Extra Dimensions  & 4D & $-4$ & $\ell$ [$\mu$m] & $\mathbf{8.6 \times 10^{9}}$ & 10--10$^3$~\cite{Johannsen:2008tm,johannsen2,Adelberger:2006dh,psaltis-RS,gnedin}\\
\hline
Time-Varying $G$  & SEP& $-4$ &  $|\dot G|$ [10$^{-12}$/yr] & $\mathbf{5.4 \times 10^{18}}$ & 0.1--1~\cite{Manchester:2015mda,2011Icar..211..401K,Hofmann,Copi:2003xd,Bambi:2005fi}\\ 
\hline \hline
\multirow{1}{*}{Massive graviton}  & \multirow{1}{*}{$m_g=0$} & \multirow{1}{*}{$+1$}  & \multirow{1}{*}{$m_g$ [eV]} & $\mathbf{10^{-22}}$~\cite{TheLIGOScientific:2016src} &  \multirow{1}{*}{$10^{-29}$--$10^{-18}$~\cite{talmadge,sutton,goldhaber,Hare:1973px,Brito:2013wya}} \\
\hline
\multirow{2}{*}{Multifractional}  & \multirow{2}{*}{LI} &  \multirow{2}{*}{$+4.75$} &  $E_*^{-1}$ [eV$^{-1}$] (time) &  \multirow{1}{*}{$\mathbf{5.8 \times 10^{-27}}$}  & ---\\ 
  &   &    &$E_*^{-1}$ [eV$^{-1}$] (space) & $\mathbf{1.0 \times 10^{-26}}$ & $3.9 \times 10^{-53}$~\cite{Kiyota:2015dla}\\ 
\hline
\multirow{2}{*}{double Special Rel.} & \multirow{2}{*}{LI} &  \multirow{2}{*}{$+5.5$} &   $\eta_{\rm dsrt}/L_{\mrm{Pl}} >0$ &  \multirow{2}{*}{$\mathbf{1.3 \times 10^{22}}$}  & ---\\ 
   &    &  &$\eta_{\rm dsrt}/L_{\mrm{Pl}} <0$  &   & $2.1 \times 10^{-7}$~\cite{Kiyota:2015dla}\\ 
\hline
\multirow{2}{*}{Extra Dimensions}    & \multirow{2}{*}{4D} &   \multirow{2}{*}{$+7$} &  $\alpha_{\rm edt}/L_{\mrm{Pl}}^2 > 0$  &  \multirow{2}{*}{$\mathbf{5.5 \times 10^{62}}$} &   $2.7 \times 10^{2}$~\cite{Kiyota:2015dla}\\ 
 &    &  & $\alpha_{\rm edt}/L_{\mrm{Pl}}^2 <0$  &  & --- \\ %\hline
\hline
 \multirow{6}{*}{Stand. Model Ext.}  & \multirow{6}{*}{LI} &   \multirow{2}{*}{$+4$} &   $\mathring{k}_{(I)}^{(4)} > 0$   &  \multirow{1}{*}{---} & $6.1 \times 10^{-17}$~\cite{Kostelecky:2010ze,Kiyota:2015dla}\\ 
 &  &    & $\mathring{k}_{(I)}^{(4)} <0$ & $\mathbf{0.64}$ &  --- \\ 
 &  &   \multirow{2}{*}{$+5.5$} &  $\mathring{k}_{(V)}^{(5)} > 0$ [cm] &  \multirow{2}{*}{$\mathbf{1.7 \times 10^{-12}}$~\cite{Kostelecky:2016kfm}}& $1.7 \times 10^{-40}$~\cite{Kostelecky:2010ze,Kiyota:2015dla}\\ 
 &    &  & $\mathring{k}_{(V)}^{(5)} <0$ [cm] & &  --- \\ 
 & &   \multirow{2}{*}{$+7$} &  $\mathring{k}_{(I)}^{(6)} > 0$ [cm$^2$] &  \multirow{2}{*}{$\mathbf{7.2 \times 10^{-4}}$} &  $3.5 \times 10^{-64}$~\cite{Kostelecky:2010ze,Kiyota:2015dla}\\ 
 &  &      & $\mathring{k}_{(I)}^{(6)} <0$ [cm$^2$] &  & --- \\ 
\hline
Ho\v rava-Lifshitz   & \multirow{1}{*}{LI} &   \multirow{1}{*}{$+7$} & \multirow{1}{*}{$\kappa_{\mrm{hl}}^4 \mu_{\mrm{hl}}^2$ [1/eV$^2$]} &  \multirow{1}{*}{$\mathbf{1.5 \times 10^{6}}$} &  \multirow{1}{*}{---}\\ 
%(\emph{Ho\v rava-Lifshitz}) &  &   & & & &\\ %\hline
\hline
Einstein-\AE ther    & \multirow{1}{*}{LI} & \multirow{1}{*}{$+4$} & \multirow{1}{*}{$c_+$} & \multirow{1}{*}{$\mathbf{0.7}$~\cite{Blas:2016qmn}} & \multirow{1}{*}{$0.03$~\cite{Yagi:2013qpa,Yagi:2013ava}}\\ 
%(\emph{Lorentz Violation}) & & & & & \\   
\noalign{\smallskip}
\hline
\hline
\end{tabular}
\end{centering}
\caption{
Various bounds on example theories that violate certain fundamental pillars in GR. The meaning of each column is as follows. 1st: names of modified theories of gravity; 2nd: GR fundamental pillars that each theory break; 3rd: the leading PN order in the gravitational waveform at which the non-GR effect enters; 4th: representative parameters in each theory; 5th: bounds on each non-GR parameter from GW150914 derived mostly in~\cite{Yunes:2016jcc}; 6th: other bounds on each theory.  Theories in the top (bottom) half of the table modifies the waveform at the level of generation (propagation). One finds similar bounds from GW151226~\cite{Yunes:2016jcc}. This table is taken and edited from~\cite{Yunes:2016jcc}. 
 }
\label{tab:summary2}
\end{table*}
\endgroup
}

Although Fig.~\ref{fig:beta-cons} is useful as the bounds presented are model independent, it is unclear what fundamental pillars of GR we are testing. Reference~\cite{Yunes:2016jcc} addressed such a question by mapping the bound on $\beta_\ppE$ to that on each example non-GR theory that violates certain fundamental aspects of GR, such as the strong equivalence principle (SEP), Lorentz invariance (LI), parity invariance (PI), 4 dimensional spacetime (4D) and a massless graviton $(m_g = 0)$. The first half of Table~\ref{tab:summary2} summarizes the results by presenting GW150914 bounds on each theory that are mapped from Fig.~\ref{fig:beta-cons}, together with other bounds from (e.g.) Solar System experiments or binary pulsar observations. We also list the fundamental pillars of GR violated by each theory, and the PN order at which dominant corrections appear. Observe first that we do not show any GW150914 bounds on the first three theories. This is because the bounds are so weak that they violate the small-coupling approximation used to derive corrections to the waveform. On the other hand, one can place meaningful constraints on the other four theories considered that break SEP, LI or the assumption of a four-dimensional spacetime. Although these GW bounds are much weaker than other existing bounds, they are the first bounds obtained in the extreme gravity regime. The potential of GW observations in probing GR is currently limited by the lack of our knowledge of non-GR effects in the merger-ringdown regime.

One can carry out a similar test for corrections generated at the level of GW propagation. For such a case, one can include corrections not only in the inspiral phase, but also in the merger-ringdown part of the waveform. GW150914 bounds on the mass of the graviton, which are three times stronger than Solar System bounds~\cite{talmadge}, were obtained in this way~\cite{TheLIGOScientific:2016src}. The second half of Table~\ref{tab:summary2} summarizes bounds on each example theory that modifies the dispersion relation of the graviton. Unlike bounds on GW generation, those on GW propagation are complementary to cosmic ray bounds obtained from the absence of the gravitational Cherenkov radiation~\cite{Kiyota:2015dla}. In other words, some of the parameter space in each theory has been constrained for the first time using GW150914. Some theories (like EA theory) are difficult to constrain from deviations in the waveform phase due to degeneracies with other parameters, like the time of coalescence. Thus the bound on such a theory in the last line of Table~\ref{tab:summary2} is obtained from bounds on the propagation speed of GWs due to the arrival time difference between the Hanford and Livingston detectors~\cite{Blas:2016qmn} (see also~\cite{Cornish:2017jml}).

Observed GW events were used to constrain other modified theories of gravity, such as non-commutative spacetime~\cite{Kobakhidze:2016cqh}. These events were also used to constrain graviton oscillations in bigravity~\cite{Max:2017flc}, a phenomenon similar to neutrino oscillations, first proposed in~\cite{DeFelice:2013nba}.

Very recently, the LVC detected yet another GW signal, that is consistent with the source being a binary neutron star merger (GW170817)~\cite{TheLIGOScientific:2017qsa}. Unlike the binary black hole events, where confirmed electromagnetic wave counterparts were absent, GW170817-associated counterpart signals were detected by gamma-rays, X-rays, ultraviolet, optical, infrared and radio waves~\cite{GBM:2017lvd}. This historic observation marks the dawn of the era of the multi-messenger astronomy.  Thanks to the simultaneous detection of GWs and electromagnetic waves, the LVC carried out new tests of GR~\cite{Monitor:2017mdv} (see also~\cite{Boran:2017rdn,Shoemaker:2017nqv,Wang:2017rpx,Wei:2017nyl}). For example, they measured the propagation speed of GWs $v_g$ from the arrival time difference between GWs and gamma-rays. Conservatively assuming that photons were emitted within 10 seconds compared to the graviton's emission time, they obtained the bound $-3 \times 10^{-15}<(v_g-c)/c<7 \times 10^{-16}$~\cite{Monitor:2017mdv}. As pointed out e.g. in~\cite{Lombriser:2015sxa,Lombriser:2016yzn}, this rules out many models in modified theories of gravity that aim to explain the current accelerating expansion of the Universe, including the quartic, quintic and covariant Galileons~\cite{Sakstein:2017xjx,Creminelli:2017sry,Ezquiaga:2017ekz,Baker:2017hug,Arai:2017hxj}. The observations constrain the parameter $c_+$ in Einstein-\AE ther theory and $\beta_\KG$ in khronometric gravity to be of order $\mathcal{O}{(10^{-15}})$~\cite{Hansen:2014ewa,Yunes:2016jcc,Baker:2017hug}, improving over previous bounds by roughly 13 orders of magnitude. Such a bound on the propagation speed of the graviton can be used to probe gravitational Lorentz violation through the SME~\cite{Kostelecky:2016kfm}. The LVC placed bounds on gravitational SME parameters which are generally a few to 10 times more stringent than previous bounds~\cite{Monitor:2017mdv}. GW170817 also probes violations of the equivalence principle. This was done by testing whether gravitons and photons feel the same gravitational potential as they propagate. The difference between the graviton's and photon's parameterized PN parameter ($\gamma_g$ and $\gamma_p$ respectively) was bounded as $-2.6 \times 10^{-7}<\gamma_g - \gamma_p< 1.2 \times 10^{-6}$~\cite{Monitor:2017mdv} via Shapiro delay measurements.
GW170817 has been used to set stringent constraints on the Vainshtein mechanism~\cite{Crisostomi:2017lbg,Langlois:2017dyl,Dima:2017pwp} and on several modifications of GR, including theories with extra dimensions~\cite{Visinelli:2017bny}, Ho\^rava gravity~\cite{Gumrukcuoglu:2017ijh}
$f(R)$ models~\cite{Nojiri:2017hai,Lee:2017dox}
and massive gravity~\cite{Heisenberg:2017qka}.

\subsubsection{Future Tests}

GW tests will benefit in the future from three types of accomplishments: (i) multiple detections, (ii) low-frequency detections, and (iii) multi-wavelength observations. It is easy to understand how multiple detections can yield improved tests: one can either stack the events to enhance the power in the signal, or simply combine the events by multiplying posteriors together to enhance constraints. This will strengthen inferences on both the generation and propagation of waves, roughly by a factor $\sqrt{N}$ if $N$ is the number of comparable signal-to-noise ratio events. In reality, the enhancement factor will be dominated by the loudest events (see e.g.~\cite{Berti:2011jz}). The second accomplishment refers specifically to observations with space-borne detectors, which will be sensitive to waves in the milli-Hz range. These observations are unique because they will allow for very high signal-to-noise ratios (in the hundreds to thousands) and very large distances (roughly Gpc and beyond), and they can probe systems at much larger separations, or much lower orbital frequency. Such observations will strengthen inferences both on the generation and propagation sectors, although propagation bounds will benefit the most, due to the long baseline of the measurements. The third accomplishment refers to observations that could be done first by space-borne detectors at deci-Hz frequencies, when the binary system is widely separated, and then again by ground-based detectors at hecto-Hz frequencies, when the \emph{same} binary merges. This will allow for precise tests that metaphorically tie the theory at both ends: during the early inspiral and during the merger simultaneously.   

\begin{figure*}[thb]
\begin{center}
\includegraphics[width=\textwidth,clip=true]{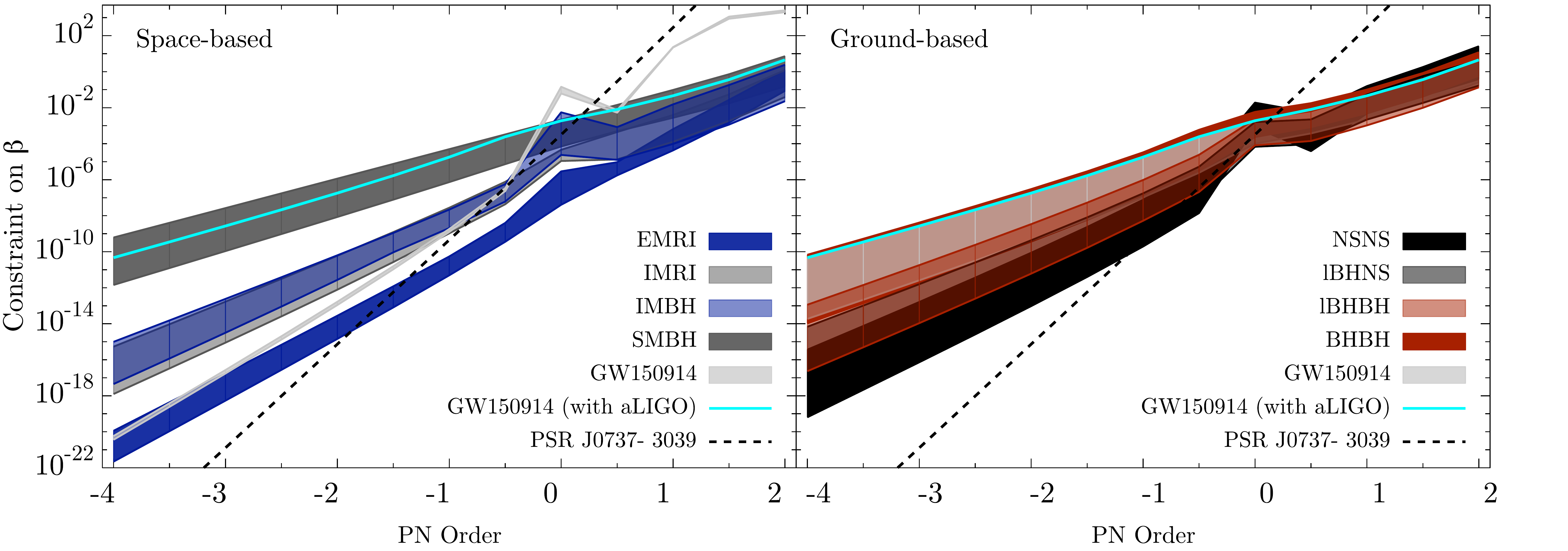}
\caption{\label{shaded} [From~\cite{Chamberlain:2017fjl}.] Projected constraints on modified gravity effects as a function of ppE PN order at which they first enter, for a variety of space-based (left) and ground-based (right) detectors and a variety of systems. Anything above the regions is projected to be ruled out. The shaded regions are bounded by the highest and lowest constraints that can be placed at a given PN order for all instruments studied. For comparison, we also include the constraints that have already been placed by aLIGO with the GW150914 detection~\cite{Abbott:2016blz,Abbott:2016nmj} (thin cyan line), as well as constraints that can be placed with binary pulsars~\cite{Yunes:2010qb} (dashed black line). Observe that the magnitudes of the projected constraints with space-based and ground-based instruments are comparable at positive PN orders, with space-based constraints being better by roughly $2$--$4$ orders of magnitude at negative PN order.}
\end{center}
\end{figure*}

Given these expected advancements, one may wonder how much more stringent future constraints will become in the future~\cite{Barausse:2016eii,Chamberlain:2017fjl,Samajdar:2017mka}.  Fig.~\ref{shaded} shows projected constraints on the $\beta_\ppE$ parameter as a function of the PN order at which they enter -- i.e., a term of $N$ PN order is associated with a ppE correction proportional to $v^{2N-5}$ -- for a variety of \emph{single} GW observations from the inspiral of compact binaries. The shaded regions correspond to variations in the constraints due to using different future instruments, including aLIGO at design sensitivity, A+, Voyager, Cosmic Explorer, the Einstein Telescope, and different incarnations of LISA.  Observe that future constraints will be many orders of magnitude more stringent than current constraints (represented here with the aLIGO observation of GW150914 in cyan). These constraints would be enhanced by a factor of roughly $\sqrt{N}$ given $N$ observations, where one should keep in mind that ground-based and space-based detectors are not expected to see the same number of events; although this is strongly dependent on the uncertain event rate, one expects to see roughly $10^{4}$ sources with ground-based instruments, and roughly $10^{2}$ sources with space-based instruments (see e.g.~\cite{Berti:2016lat}).

\begin{figure*}[thb]
\begin{center}
\includegraphics[width=8.5cm,clip=true]{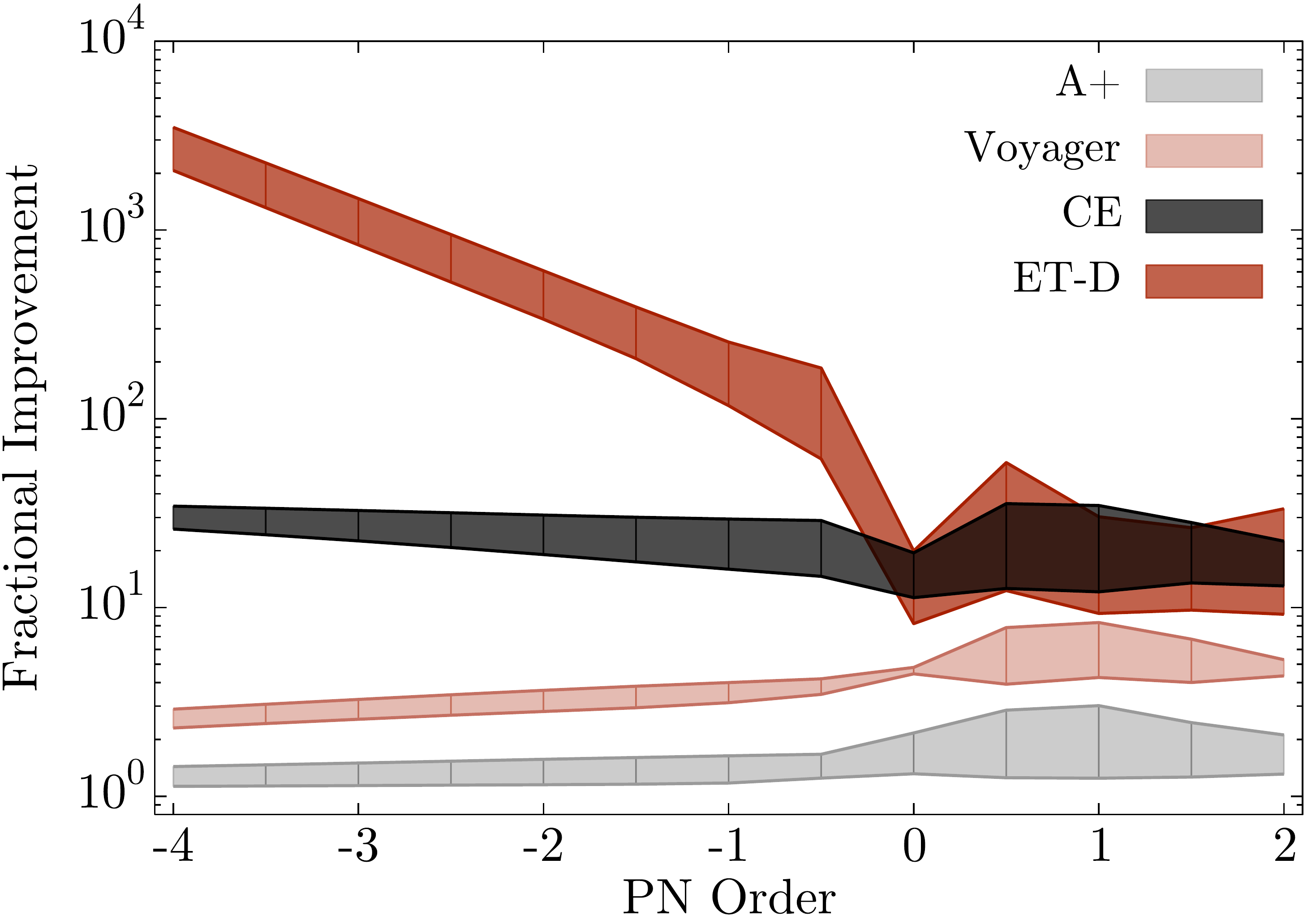}
\caption{\label{frac-improv} [From~\cite{Chamberlain:2017fjl}.] Projected fractional improvement of constraints on modified gravity effects as a function of ppE PN order at which they first enter for a variety of detectors (relative to aLIGO at design sensitivity). The shaded regions are the same as in Fig.~\ref{shaded}. Observe that 3G detectors improve constraints by more than an order of magnitude even for single detectors.}
\end{center}
\end{figure*}

A related question is how the strength of the constraint changes with different third-generation (3G) detector configuration. Fig.~\ref{frac-improv} shows the fractional improvement of projected constraints as a function of ppE PN order at which modified gravity effects first enter~\cite{Chamberlain:2017fjl} for a variety of detectors (relative to aLIGO at design sensitivity). Although minor upgrades, like A+ and Voyager, will only lead to modest improvements in constraints, 3G detectors can achieve improvements that are better than an order of magnitude. The fractional improvement dramatically increases at negative PN order in the ET case, simply because of this detector's greatly improved sensitivity at low frequencies~\cite{Chamberlain:2017fjl}. 

\begin{figure*}[htb]
\begin{center}
\includegraphics[width=5.75cm,clip=true]{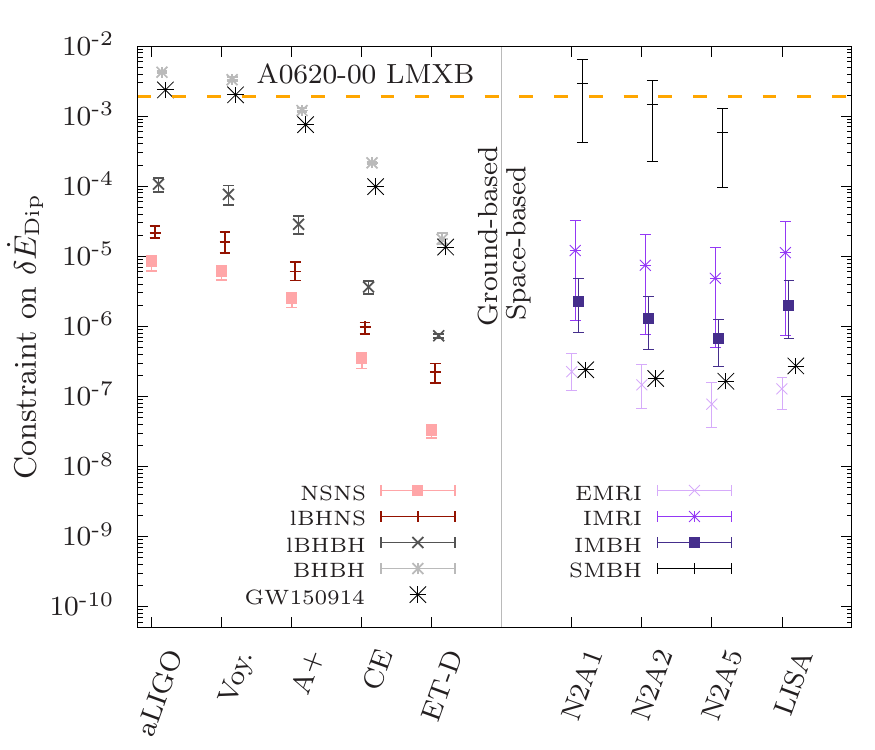}
\includegraphics[width=5.5cm,clip=true]{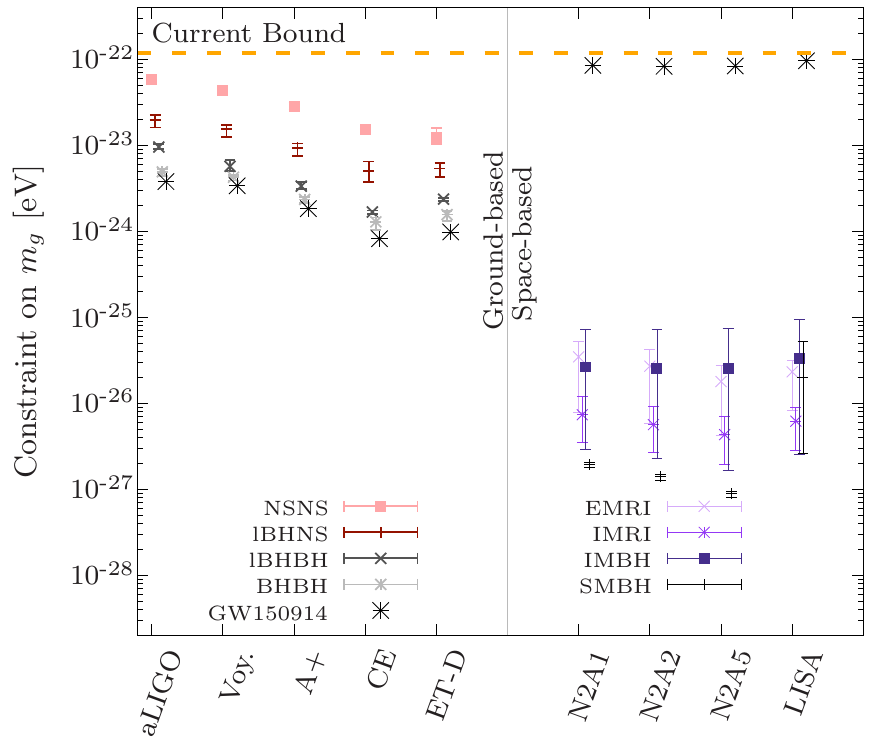}
\caption{\label{cons} [From~\cite{Chamberlain:2017fjl}.] Projected constraints on dipole emission (left) and the mass of the graviton (right) for a variety of sources and detectors. Anything above the points is projected to be ruled out. The horizontal dashed lines correspond to current constraints. Observe that the magnitude of the projected constraints with space-based and ground-based instruments are always a great improvement over what aLIGO at design sensitivity will be able to achieve.}
\end{center}
\end{figure*}

Given these more stringent constraints, the natural question to ask
is: what new physics will be probed in the future? Multi-wavelength
observations with space- and ground-based instruments will allow for
constraints on violations of the strong equivalence principle that are
8 orders of magnitude more stringent than all current
bounds~\cite{Barausse:2016eii}. Single observations with future
instruments will allow for constraints on the size of a large
extra-dimension (in Randall-Sundrum type models) that are 5 orders of
magnitude more stringent than current bounds with
aLIGO~\cite{Chamberlain:2017fjl}.  Similarly, constraints on the mass
of the graviton from propagation effects in the dispersion relation
will be about 5 orders of magnitude better than current
bounds~\cite{Chamberlain:2017fjl}. These constraints would begin to
approach the natural value of the mass of the graviton in eV that one
would expect if such a mass is somehow connected to a solution to the
dark-energy problem. Fig.~\ref{cons} shows how constraints on dipole
emission and the mass of the graviton improve for a variety of sources
and detectors~\cite{Chamberlain:2017fjl}.

%%%%%%%%%%%%%%%%%%%%%%%%%%%%%%%
\section{Merger Tests of Modified Gravity}
\label{sec:merger}

There are very few merger simulations in non-GR theories. This is
because simulations are possible only for theories where a gauge has
been found in which the theory is well-posed. For example, dCS gravity
is likely to be ill-posed when treated as an exact
theory~\cite{Delsate:2014hba,Crisostomi:2017ugk}, although it becomes
well-posed if one treats it as an effective field theory and solves
the field equations order by order~\cite{Delsate:2014hba}, as done
in~\cite{Okounkova:2017yby}. The well-posedness of Lovelock gravity
and of a certain subclass of Horndeski theories has been studied
in~\cite{Papallo:2017qvl}, including also D$^2$GB
gravity~\cite{Papallo:2017ddx}. The latter theory was found to be not
strongly hyperbolic in a \emph{specific} class of (generalized
harmonic) gauges.  Therefore the well-posedness of EdGB and D$^2$GB
gravity in general is still an open question. Recently Cayuso
\textit{et al}.~\cite{Cayuso:2017iqc} proposed a new framework to make
a truncated theory well-behaved following the Israel-Stuart formalism
for relativistic viscous hydrodynamics, and they applied this
framework to toy models for linearized non-commutative geometry and
dCS gravity. The application of this framework to the full theories
and to other non-GR theories is an active research area.

In contrast to the theories discussed above, proving well-posedness is
almost trivial for the Bergmann-Wagoner scalar-tensor
theories\footnote{Since $f(R)$ theories are equivalent to
  scalar-tensor gravity, they are also
  well-posed~\cite{LanahanTremblay:2007sg,Paschalidis:2011ww}.}
discussed in
section~\ref{sec:BergmannWagoner}~\cite{Salgado:2008xh}. We will first
review analytical work on the dynamics of binary systems in these
theories, and then summarize some numerical simulations of compact
binary systems that have been carried out in these theories.

% ------------------------------------------------------------------
\subsection{Compact Binaries in Scalar-Tensor Theories: Analytical Results}

As stated earlier, scalar-tensor gravity is the most studied modified
theory of gravity. It was studied in depth already in the
1950s~\cite{Brans:1961sx,Fierz:1956zz,Jordan:1959eg}, and most
variants of Bergmann-Wagoner theories have been constrained to be
extremely close to GR, at least in the weak-field limit. These
theories are well-posed, so compact binary systems can be evolved
throughout merger. The question is: are there variants of
scalar-tensor gravity which are still compatible with weak-field and
cosmological observations, but at the same time make predictions in
the strong-gravity regime that are sufficiently different from GR to
be testable with GW interferometers?

Weak-field observations imply that $\alpha_0\ll1$ in
Eq.~\eqref{DEalphabeta}, so that deviations from GR are generally
small. However there are cases where scalar-tensor gravity may lead to
observable differences from GR in the strong-field regime targeted by
GW detectors. Three such smoking guns of scalar-tensor gravity are:

{\em (i) The emission of dipolar gravitational radiation from compact
  binary systems}. Dipolar gravitational radiation is
``pre-Newtonian,'' i.e.~it occurs at lower PN order than quadrupole
radiation, and it does not exist in
GR~\cite{1975ApJ...196L..59E,Will:1989sk}. So far, LIGO/Virgo binary
black hole detections outnumber the detections of binary systems
containing neutron stars. Unfortunately, the phenomenology of
scalar-tensor theory in vacuum spacetimes (such as binary black hole
spacetimes) is less interesting than in the presence of matter. This
is because, when the matter action $S_M$ can be neglected, the
Einstein-frame formulation of the theory is equivalent to GR minimally
coupled to a scalar field. Isolated black holes in Bergmann-Wagoner
theories\footnote{Some classes of Horndeski theory (for example, those
  that can be shown to be equivalent to Einstein-dilaton-Gauss-Bonnet
  gravity through integration by parts) are such that these no-hair
  theorem can be
  circumvented~\cite{Sotiriou:2013qea,Sotiriou:2014pfa,Maselli:2015yva,Silva:2017uqg},
  so that stationary black hole solutions can be different from GR.}
satisfy the same {\it no-hair theorem} as in GR, and thus the
stationary black hole solutions in the two theories
coincide\footnote{There are some proposal to circumvent these no-hair
  theorems involving time-dependent scalar
  fields~\cite{Herdeiro:2014goa,Herdeiro:2015gia,Herdeiro:2015waa}.
  Recent evidence shows that the resulting solutions are
  unstable~\cite{Ganchev:2017uuo}, but the instability is
  astrophysically irrelevant in some regions of the parameter
  space~\cite{Degollado:2018ypf}.}~\cite{Heusler:1995qj,Sotiriou:2011dz}. Dipolar
radiation is proportional to the difference in sensitivity between the
two binary members [see Eq.~\eqref{sensitivity} below], so it vanishes
identically for binary black hole systems. In fact, as we explain in
Appendix~\ref{app:pn_st}, dynamical (vacuum) black hole binary
spacetimes satisfy what we could call a {\it generalized no-hair
  theorem}: the dynamics of a black hole binary system in
Bergmann-Wagoner theory with vanishing potential in asymptotically
flat spacetimes are the same as in GR up to at least $2.5$ PN order
for generic mass ratios~\cite{Damour:1992we,Mirshekari:2013vb}, and at
any PN order in the extreme mass-ratio limit~\cite{Yunes:2011aa}.

{\em (ii) Spontaneous scalarization.} Sizable strong-field
deviations from GR can occur in the presence of non-perturbative neutron star (or
black hole) solutions for which the scalar field amplitude is finite even when
$\alpha_0\ll1$. This {\it spontaneous scalarization}
phenomenon~\cite{Damour:1993hw,Damour:1996ke} could significantly
affect the mass and radius of a neutron star, and therefore the orbital motion
of a compact binary system, even far from coalescence. This mechanism
relies on the presence of matter, so (once again) it does not apply to
black holes; however, it has recently been pointed out that there are some
scalar-tensor theories where black hole solutions can
scalarize~\cite{Doneva:2017bvd,Silva:2017uqg,Antoniou:2017acq,Antoniou:2017hxj}.

{\em (iii) Superradiance in rotating black hole spacetimes.}
Another class of non-perturbative mechanisms involves rotating black holes and
the phenomenon known as superradiance (see~\cite{Brito:2015oca} for a review). The coupling of
massive scalar fields to matter in orbit around rotating black holes leads to
a surprising effect: because of superradiance, matter could in
principle hover into ``floating orbits'' for which the net
gravitational energy loss at infinity is entirely provided by the black hole's
rotational energy~\cite{Cardoso:2011xi}.  This phenomenon could in
principle occur in nature, but it is possible that radiation
reaction~\cite{Zimmerman:2015hua} and finite-size
effects~\cite{Fujita:2016yav} will destabilize floating
orbits. Tachyonic instabilities similar to spontaneous scalarization
can also be produced when ``ordinary'' rotating black holes in GR are
surrounded by matter, as long as the trace of the stress-energy tensor
changes sign~\cite{Cardoso:2013opa,Cardoso:2013fwa}.

Two complementary approaches are used to model these effects in compact binary systems: analytical calculations (usually PN expansions) and numerical relativity simulations. Below we will review {\em numerical studies} of compact binaries in scalar-tensor gravity and other modified theories of gravity.

%%%%%%%%%%%%%%%%%%%%%%%%%%%%%%%%%%%%%%%%%
\subsection{Compact Binaries in Scalar-Tensor Theories: Numerical Simulations}\label{sec:nr_st}
%%%%%%%%%%%%%%%%%%%%%%%%%%%%%%%%%%%%%%%%%

Even within GR, obtaining numerically stable and accurate time
evolutions of the Einstein equations took almost 50 years. A stable
evolution of binary black holes and neutron stars in numerical relativity requires an
understanding of many complex issues, such as the well-posedness of
the evolution system, the construction of initial data and gauge
conditions~\cite{Baumgarte:2010ndz}.  These same questions naturally
arise also in modified theories of gravity, and at present they remain
unanswered for most of these theories. Scalar-tensor theories are a
notable exception, because they can be formulated in close analogy to
GR.  As discussed in Section~\ref{sec:ST}, the action of scalar-tensor
theories in the Einstein frame is the same as the Einstein-Hilbert
action, except for a minimal coupling with the scalar field in the
gravitational sector. A non-minimal coupling with the scalar field only
appears in the matter sector.  The resulting field equations are
similar to the field equations of GR, and the evolution of the scalar
field $\varphi$ is dictated by a hyperbolic wave equation. Therefore
scalar-tensor theories can be evolved using relatively minor
generalizations of the numerical codes developed for GR. As shown by
Salgado et al.~\cite{Salgado:2005hx,Salgado:2008xh}, a strongly
hyperbolic formulation can be obtained also in the physical (Jordan)
frame. However, the Einstein frame is exceptionally convenient for
applications to binary black hole spacetimes, because in vacuum the evolution
equations are independent of the coupling function $A(\varphi)$. In
this sense, a single numerical evolution represents a whole class of
theories characterized by different functional forms of $A(\varphi)$
for a given potential $V(\varphi)$.  Different choices of the function
$A(\varphi)$ result in different physical predictions, but all of
these predictions can be calculated by {\em post-processing} data from
a single numerical simulation. This would not be possible in the
Jordan frame, where the coupling function appears explicitly in the
field equations.

%=============================================================================
\paragraph{Black hole binaries.}
\label{sec:NRblackholes}

Because of the ``dynamical no-hair theorem'' summarized in
Appendix ~\ref{app:pn_st}, interesting black hole binary dynamics that are
significantly different from GR requires somewhat contrived scenarios.
The ``dynamical no-hair theorem'' of Appendix~\ref{app:pn_st} relies on the following
assumptions: (1) vacuum spacetime, (2) the scalar-tensor action is
truncated at second order in a derivative expansion, (3) the potential
$V(\varphi)$ vanishes, and (4) the metric is asymptotically flat, with
an asymptotically constant scalar field.

Deviations from GR in black hole binaries can occur if we violate any of these
assumptions.  What happens when we violate hypothesis (1) will be
discussed in the next section. Violating hypothesis (2) by introducing
higher-order derivatives in the action would lead to substantially
more complicated equations, whose well-posedness is presently unclear
(see e.g.~\cite{Berti:2013gfa}).

Healy et al.~\cite{Healy:2011ef} introduced nontrivial dynamics by
placing the black holes inside a scalar field ``bubble'' which in some cases
includes a non-vanishing scalar field potential, thus violating
hypothesis (3). As the bubble collapses, the black holes accrete the scalar
field and grow in mass. This mass growth affects the binary dynamics
and the emitted gravitational radiation, and the inclusion of a
potential term introduces longer lived dynamics in the scalar field
mode. Their work supports the view that an evolving scalar field is
required to introduce interesting dynamics in black hole binaries. However,
for the effects to be observable the merging black holes must accrete enough
scalar field to appreciably change their masses and modify the binary
evolution.

Horbatsch \& Burgess~\cite{Horbatsch:2011ye} suggested violating
hypothesis (4) instead: if the scalar field is time-dependent at the
boundary, the black holes in a binary could retain scalar
hair~\cite{Jacobson:1999vr} and emit dipole radiation, as long as
their masses are not exactly equal. Numerical evolutions implementing
this idea were carried out in~\cite{Berti:2013gfa} by introducing
non-asymptotically flat or constant boundary conditions. The main
motivation for relaxing this assumption comes from cosmology: indeed,
inhomogeneous scalar fields have been considered as an alternative to
dark matter~\cite{Sahni:1999qe,Hu:2000ke} and as models of
supermassive boson stars~\cite{Macedo:2013qea}.  For scalar-field
profiles that vary on a length scale much larger than the black hole
binary orbital separation, the scalar-field gradient can be considered
approximately constant.  Ref.~\cite{Berti:2013gfa} studied the
quasi-circular inspiral of a non-spinning black hole binary (with mass
ratio $3:1$) in a scalar-field gradient perpendicular to the orbital
angular momentum vector.  The lowest multipoles of the radiation are
effectively indistinguishable from their GR counterparts. However, a
non-vanishing scalar field gradient can lead to (mostly dipolar)
emission of scalar radiation, which is not present in GR at {\em
  twice} the orbital frequency. At first glance this may appear
surprising, but a simple calculation reveals that this feature is a
consequence of the interaction of the orbital motion with a background
field with an azimuthal component $m=1$: cf.~the discussion
around~Eqs.~(36)-(38) of~\cite{Berti:2013gfa}.

In summary, the simulations of~\cite{Berti:2013gfa} and~\cite{Healy:2011ef}
demonstrate that non-asymptotically flat
boundary conditions provide a mechanism to generate scalar radiation
in black hole inspirals in scalar-tensor theories of gravity, {\em at least in
  principle}. Unfortunately, this scalar radiation will be
unobservable in the near future for cosmologically realistic values of
the scalar-field gradients.

%=============================================================================
\paragraph{Neutron star binaries.}
\label{sec:NRneutronstars}

The dynamics of scalar-tensor theories of gravity in the presence of
matter sources can be very different from GR. The theory can violate
the strong equivalence principle, so that self-gravitating objects
follow trajectories that depend on their internal
composition/structure: this is the well-known ``Nordtvedt
effect''~\cite{Nordtvedt:1968qr,Roll:1964rd,1975ApJ...196L..59E}.

For Bergmann-Wagoner theories, the dimensionless coupling
$\alpha(\varphi)$ between the scalar field and matter can be expanded
as in Eq.~(\ref{DEalphabeta}), where
$\alpha_0=1/\sqrt{3+2 \omega_{BD}}$ and $\beta_0$ are dimensionless
constants, and $\varphi_0$ is the asymptotic value of the scalar
field.  The leading-order term in this expansion, i.e.  $\alpha_0$, is
severely constrained by Solar System experiments:
$\omega_{BD} > 40\,000$, or
$\alpha_0<3.5\times 10^{-3}$~\cite{Will:2014kxa}. Furthermore
${\beta_0}\gtrsim-4.5$, otherwise spontaneous scalarization would
affect the dynamics in ways that are severely constrained by binary
pulsar data (note that these constraints are mildly dependent on, and
slightly degenerate with, the equation of state of nuclear
matter~\cite{Shibata:2013pra,Silva:2014fca,Shao:2017gwu}).

Recently, Barausse et al.~\cite{Barausse:2012da} and Palenzuela et
al.~\cite{Palenzuela:2013hsa} (using numerical simulations and
semi-analytical arguments, respectively) discovered a phenomenon
similar to spontaneous scalarization in the late stages of the
evolution of binary neutron star binaries, that they called ``dynamical
scalarization'' (see also~\cite{Shibata:2013pra,Sennett:2016rwa}).
Even when the individual neutron stars would \textit{not} spontaneously
scalarize in isolation, the scalar field inside each star can grow in
amplitude when the binary separation decreases to about
$50-60~{\rm km}$, affecting the binary dynamics and speeding up the
merger with respect to GR.  The resulting gravitational waveforms are
significantly different from GR at frequencies
$\sim 500-600~{\rm Hz}$, and deviations at even lower frequencies are
possible for certain binary systems and theory parameters.  Therefore,
the effects of dynamical scalarization are in principle detectable (at
least in some cases) by LIGO/Virgo like detectors for values of the
coupling parameters $\omega_0$ and $\beta$ that are {\em still
  allowed} by Solar System and binary pulsar
tests~\cite{Shao:2017gwu}: see~\cite{Sampson:2014qqa} for an extensive
discussion.  

Unfortunately, the coupling parameters $(\alpha_{0},\beta_{0})$ needed
in order to obtain any type of scalarization seem to be incompatible
with Solar System observations when accounting for the cosmological
evolution of the scalar
field~\cite{Damour:1992kf,Damour:1993id,Sampson:2014qqa,Anderson:2016aoi}. This
is because for scalarization to occur, typically $\beta_{0}$ must be
negative (or positive and very large~\cite{Mendes:2016fby}, which
leads to instabilities~\cite{Palenzuela:2015ima}), but this also
forces the scalar field to evolve cosmologically away from General
Relativity, thus leading to maximal deviations by the present epoch in
the Solar System. That is, when accounting for the cosmological
evolution of the scalar field, the observation of the Shapiro time
delay and its consistency with GR essentially require $\beta_{0}$ to
be positive (and scalarization not to occur) in the simplest
models. One way out of this is to allow the scalar field to have a
mass or to fine-tune the cosmological initial conditions, so that the
scalar field does not evolve cosmologically at all.

Deviations from GR may also be observable in the
electromagnetic signal from binaries of magnetized
NSs~\cite{Ponce:2014hha}, and therefore they could be tested by
multi-messenger observations similar to
GW170817~\cite{TheLIGOScientific:2017qsa}. This work was recently
generalized to certain $f(R)$ models -- namely, $f(R)\sim R^2$ --
which are equivalent to massive scalar-tensor theories. The
inspiral/merger dynamics of neutron star binaries in these theories, as observed
by GW detectors, can set stringent bounds on attractive finite-range
scalar forces~\cite{Sagunski:2017nzb}.

%------------------------------------------------------------------
\subsection{Compact binaries in other theories of gravity}

As discussed above, our ability to simulate compact binary mergers in
modified theories of gravity that go beyond ``plain vanilla''
scalar-tensor theories is limited by our poor understanding of their
well-posedness.

Jai-akson et al.~\cite{Jai-akson:2017ldo} estimated the outcome of a
merger in Einstein-Maxwell-dilaton gravity (whose action arises in the
low-energy limit of string theory) using a ``geodesic analogy''
approach inspired by~\cite{Buonanno:2007sv}, rather than numerical
simulations. Their qualitative conclusions were confirmed by
full-blown numerical simulations of isolated and binary black hole
systems~\cite{Hirschmann:2017psw}. For the coupling parameters
considered in their work, the dilaton can largely be ignored, and GW
signals from binary black hole systems in these theories are hard to
distinguish from their GR counterparts.

Another line of inquiry tries to understand (or bypass) well-posedness
issues, such as the study of the initial value problem in Lovelock
gravity and in a subclass of Horndeski theories~\cite{Papallo:2017qvl}
including EdGB gravity~\cite{Papallo:2017ddx}, and the
Israel-Stewart~\cite{Israel:1976tn,Israel:1979wp} type
approach~\cite{Cayuso:2017iqc}, as we already reviewed.  Another idea
to cure these pathologies involves an iterative strategy, where the
zeroth-order solution provided by GR is employed to evaluate
higher-order ``corrections''~\cite{Endlich:2017tqa}.  Okounkova et
al.~\cite{Okounkova:2017yby} performed binary black hole merger
simulations in dCS gravity using a well-posed perturbation scheme for
numerically integrating beyond-GR theories that have a continuous
limit to GR, and working to linear order in the perturbation
parameter. They computed gravitational waveforms in GR and energy
fluxes of the dCS pseudo-scalar field, finding good agreement with
analytic predictions at early times~\cite{Yagi:2011xp} (including the
absence of pseudo-scalar dipole radiation) and discovering new
phenomenology, including in particular a burst of dipole radiation
during merger. Perhaps unsurprisingly, they found that LIGO
observations could place bounds on the new dCS length scale that are
approximately comparable to the size of the black hole horizon,
i.e. $\sim {\cal O}(10)$~km.

More interesting merger dynamics could occur in the presence of black
hole scalarization phenomena, which have recently been studied in
Einstein-Maxwell-dilaton~\cite{Julie:2017rpw} and
scalar-Gauss-Bonnet~\cite{Silva:2017uqg} gravity models.

%%%%%%%%%%%%%%%%%%%%%%%%%%%%%%%
\section{Outlook and Discussion}
\label{sec:outlook}

We end this review by discussing some important future directions that
need to be pursued. In order to improve our ability to probe extreme
gravity with GWs, one needs to prepare parametrized template waveforms
that capture non-GR effects not only in the inspiral, but also in the
merger-ringdown phase. Ideally one would achieve this goal through black hole
merger simulations in various non-GR theories, which are necessary to
understand qualitatively and quantitatively how non-GR modifications
affect the merger-ringdown phase, and to construct complete
inspiral-merger-ringdown parameterized waveforms. These simulations
require a preliminary understanding of the well-posedness properties
of modified theories of gravity. Some recent attempts to build
parametrized frameworks for the post-merger/ringdown waveform (see
e.g.~\cite{Glampedakis:2017dvb,Tattersall:2017erk}) will be discussed
in the second part of this contribution.

Current work linking GW observations with the ``fundamental''
parameters of various modified theories of
gravity~\cite{Yunes:2016jcc} can and should be improved in various
ways. First, one needs to use the actual data and carry out a Bayesian
analysis to derive bounds on modified theories of gravity with GW
observations. Second, one needs to study how the bounds on each theory
improve by combining all the observed GW events. Bounds on generic
non-GR parameters in the waveform with multiple events have been
derived by the LIGO/Virgo Collaboration
in~\cite{TheLIGOScientific:2016pea,Abbott:2017vtc,Abbott:2017oio},
though it is difficult to map such combined bounds on generic
parameters to those on theoretical coupling constants in each
theory. This is because the former depends not only on the latter, but
also on source parameters (like masses) that are different from one
source to another. Thus, to derive combined bounds on each theory, one
may need to carry out a theory-based (rather than model-independent)
analysis. Third, one needs to repeat and extend these calculations
taking into account the binary neutron star merger event GW170817.

Other ways to improve bounds on non-GR theories include deriving
subleading PN corrections to the gravitational waveform. Higher-order
corrections, while subdominant in certain theories (such as
Brans-Dicke theory~\cite{Yunes:2016jcc}), may be important in other
theories. For example, the leading $-1$PN correction in EdGB gravity
is suppressed for nearly equal-mass and equal-spin black hole binaries
(unless scalarization occurs~\cite{Silva:2017uqg}), and thus the
subleading 0PN correction may dominate the $-1$PN contribution.

Another avenue for future work includes deriving black hole
sensitivities in theories like EA and khronometric gravity. These
sensitivities can be calculated by constructing a slowly moving black
hole solution relative to the vector field, following previous work on
neutron star
sensitivities~\cite{Yagi:2013qpa,Yagi:2013ava}. Reference~\cite{Yunes:2016jcc}
used the 0PN correction to the waveform for these theories as the
dominant contribution, but the $-1$PN correction due to scalar and
vector radiation (that depends on the sensitivities) may dominate over
the 0PN correction.

One also needs to construct non-GR waveforms for generic binaries with
precessing and eccentric orbits. For example, the dCS correction to
the waveform in Table~\ref{tab:mapping-ppE} is only valid for
spin-aligned systems, and thus it is important to derive the waveform
for precessing binaries in this theory, as spins are crucial to
constrain parity violation in gravity. However in order to construct
parameterized non-GR template waveforms for generic orbits, one needs
to first construct such waveforms with sufficient reliability {\em
  within GR}.

Another important avenue to pursue is to reveal the importance of
corrections to GW generation when probing corrections to GW
propagation. Reference~\cite{Yunes:2016jcc} showed that generation
effects are negligible compared to the propagation effects for a
specific model of massive gravity, but this may not be true in other
modified gravity theories.

\begin{acknowledgements}
E.B. was supported by NSF Grants No. PHY-1607130 and AST-1716715.
N.Y. acknowledges support through the NSF CAREER Grant PHY-1250636 and NASA Grants NNX16AB98G and 80NSSC17M0041.
\end{acknowledgements}

\appendix

\section{Derivation of the Black Hole Scalar Charge in decoupled dynamical Gauss-Bonnet Gravity}
\label{app:scalar-charge}

The goal of this appendix is to derive the scalar charge in D$^2$GB gravity for a stationary black hole, valid to arbitrary order in spin. 
We closely follow the calculation in dCS gravity in~\cite{Yagi:2012vf}.
The scalar charge $\mu$ 
can be read off from the $1/r$ coefficient in 
the asymptotic behavior of the scalar field at spatial infinity as $\phi = \mu\, M/r + \mathcal{O}(M^2/r^2)$, where $M$ is the black hole mass.
Since we work within 
the small-coupling approximation, we can take the background metric to be Kerr, and 
the above equation becomes
\be
\frac{\partial}{\partial \tilde r} \left( \Delta \frac{\partial \phi}{\partial \tilde r} \right) + \frac{1}{\sin \theta} \frac{\partial}{\partial \theta} \left( \sin\theta \frac{\partial \phi}{\partial \theta} \right) =T\,,
\label{scalar-eq}
\ee
where we work in the rescaled radial coordinate $\tilde r \equiv r/M$ and $\Delta \equiv \tilde r^2 - 2M \tilde r + \chi^2$ with $\chi$ representing the dimensionless Kerr parameter, 
and
\ba
T &\equiv & - 48 \frac{\alpha_\GB\, M^2}{\Sigma^5} \left[\tilde r^6 -15 \tilde r^4 \chi^2
   \cos ^2 \theta 
%\right. \nn 
%\\
%   & & 
%\left. 
+15 \tilde r^2 \chi^4
   \cos ^4 \theta
    - \chi^6 \cos
   ^6 \theta  \right]\,,
\ea
with  $\Sigma \equiv \tilde r^2 + \chi^2 \cos^2\theta$.

In order to solve the above field equation using Green's functions, we decompose the scalar field $\phi$ and 
the source term $T$ as~\cite{Teukolsky:1973ap}
\ba
\label{eq:scalar-EdGB-decomp}
\phi &=& \frac{\alpha_\GB}{M^2} \sum_{\ell}   R_{\ell}(\tilde r)\, S_{\ell}(\theta)\,, \\
T &=&  \frac{\alpha_\GB}{M^2} \sum_{\ell} T_{\ell}(\tilde r)\, S_{\ell}(\theta)\,,
\label{T-decomp}
\ea
where $S_\ell (\theta)$ is normalized as
\be
2 \pi \int_0^\pi S_\ell^2\, \sin\theta\, d\theta = 1\,.
\ee
Inverting Eq.~\eqref{T-decomp}, one obtains
\be
T_\ell = 2 \pi \frac{M^2}{\alpha_\GB} \int_{0}^{\pi} T\, S_{\ell}\, \sin \theta\, d\theta\,.  
\ee
Eq.~\eqref{scalar-eq} can be split into radial and angular parts as
\ba
\label{eq:scalar-EdGB-radial}
\frac{\partial}{\partial \tilde r} \left( \Delta \frac{\partial R_{\ell}}{\partial \tilde r} \right) -\ell (\ell +1) R_{\ell} &=& T_{\ell}\,, \\ 
\frac{1}{\sin \theta} \frac{\partial}{\partial \theta} \left( \sin\theta \frac{\partial S_{\ell}}{\partial \theta} \right) + \ell (\ell +1) S_{\ell} &=& 0\,.
\ea
The solution to the second equation is nothing but the $m=0$ mode of the spherical harmonics $S_{\ell} = Y_{\ell 0}$.

Let us first derive the scalar \emph{monopole} charge by concentrating on the $\ell = 0$ mode.
The solution to Eq.~\eqref{eq:scalar-EdGB-radial} consists of homogeneous and particular solutions. 
Let us first study the former.
Modulo overall integration constants, homogeneous solutions for the $\ell = 0$ mode
of Eq.~\eqref{eq:scalar-EdGB-radial} are given by
\ba
R_0^{(\mathrm{hom},1)} (\tilde r) &=& 1\,, \\
R_0^{(\mathrm{hom},2)} (\tilde r) &=& \frac{1}{2 \sqrt{1-\chi
   ^2}} \log
   \left(\frac{\tilde r-1-\sqrt{1-\chi
   ^2}}{\tilde r-1 + \sqrt{1-\chi
   ^2}}\right)\,. \nn \\
\ea
The asymptotic behavior of $R_0^{(\mathrm{hom},2)}$ at spatial infinity and at 
the Kerr horizon $\tilde{r}_\mathrm{hor} \equiv 1 + \sqrt{1-\chi^2}$ is given by
\ba
R_0^{(\mathrm{hom},2,\mathrm{inf})} (\tilde r) &=& \frac{1}{\tilde{r}} + \mathcal{O}\left( \frac{1}{\tilde{r}^{2}} \right)\,,  \\
R_0^{(\mathrm{hom},2,\mathrm{hor})} (\tilde r) &=& \frac{1}{2 \sqrt{1-\chi^2}} \log \left( \frac{\tilde{r}-\tilde{r}_\mathrm{hor}}{2\sqrt{1-\chi^2}}   \right)+ \mathcal{O}[(\tilde{r}-\tilde{r}_\mathrm{hor})]\,.
\ea
Since EdGB gravity is a shift-symmetric theory, one can set $\phi(\infty) = 0$ without loss of generality (namely no contribution from $R_0^{(\mathrm{hom},1)}$). Imposing further regularity at the horizon, one finds that the homogeneous solution is absent.

Let us now turn our attention to the particular solution $R_0^{(\mathrm{p})} (\tilde r)$. 
Such a solution is obtained by using the Green's function constructed from the two 
independent homogeneous solutions above~\cite{Mino:1997bx,Hughes:1999bq,Sasaki:2003xr}:
\ba
\label{eq:scalar-EdGB-particular}
R_0^{(\mathrm{p})} (\tilde r) &=& \frac{1}{\Delta \; W} \left[ R_0^{(\mathrm{hom},2)} (\tilde r) \int_{\tilde r_\mathrm{hor}}^{\tilde r} T_0(\tilde r')\, R_0^{(\mathrm{hom},1)} (\tilde r')\,  d\tilde r' \right. \nn \\
& & \left. - R_0^{(\mathrm{hom},1)} (\tilde r) \int_\infty^{\tilde r} T_0 (\tilde r') \,R_0^{(\mathrm{hom},2)} (\tilde r') \, d\tilde r' \right]\,,   
\label{particular}
\ea
where $W$ is the Wronskian:
\be
W  \equiv R_0^{(\mathrm{hom},1)}\, \frac{d}{d\tilde r} R_0^{(\mathrm{hom},2)} - R_0^{(\mathrm{hom},2)}\, \frac{d}{d\tilde r} R_0^{(\mathrm{hom},1)} 
= \frac{1}{\Delta}\,. 
\ee
The lower bound of the integral in Eq.~\eqref{particular} is determined such that the solution is regular at the horizon and satisfies $\phi(\infty) = 0$. 
For the purpose of studying the leading asymptotic behavior at infinity, one only needs to
consider the first term in Eq.~\eqref{eq:scalar-EdGB-particular}.

Combining Eqs.~\eqref{eq:scalar-EdGB-decomp} and~\eqref{eq:scalar-EdGB-particular} and performing the integral in the latter, one reads off the monopole scalar charge as
\ba
   \label{eq:scalar-charge-D2GB}
\mu^\GB &=& \frac{\alpha_\GB}{M^2} \frac{Y_{00}}{\Delta \; W} \int_{\tilde r_\mathrm{hor}}^\infty T_0(\tilde r')\, R_0^{(\mathrm{hom},1)} (\tilde r')\, d\tilde r' \nn \\
& = & 4 \frac{\alpha_\GB}{M^2} \frac{\sqrt{1-\chi
   ^2}-1 + \chi ^2}{\chi ^2}\,.
\ea
We checked that when we expand the above scalar charge around $\chi = 0$, 
the expression agrees with that in~\cite{Ayzenberg:2014aka} to $\mathcal{O}(\chi^8)$.

In a similar manner, one can calculate the \emph{quadrupolar} scalar charge $q^\GB$ by extracting the coefficient of $P_2(\cos \theta) M^3/r^3$ in the asymptotic behavior of the scalar field at spatial infinity. One finds 
\ba
   q^\GB &=& -\frac{4}{3 \chi ^3} \frac{\alpha}{M^2} \left\{\chi  \left[2 \chi ^2 \left(\chi
   ^2+\sqrt{1-\chi ^2}-2\right)  -5
   \sqrt{1-\chi ^2}+8\right]+6 \tan
   ^{-1}\left(\frac{\sqrt{1-\chi ^2}-1}{\chi
   }\right)\right\}\,. \nn \\
\ea
Again, we checked that an expansion of this expression about $\chi=0$ agrees with that in~\cite{Ayzenberg:2014aka} to $\mathcal{O}(\chi^8)$.

%------------------------------------------------------------------------------------
\section{A ``dynamical no-hair theorem'' for black holes in scalar-tensor gravity}
\label{app:pn_st}

The goal of this appendix is to show and explain how black hole binaries do not develop
scalar hair upon dynamical evolution. That is, we will explain how the
dynamics of a black hole binary system in Bergmann-Wagoner theory with
vanishing potential in asymptotically flat spacetimes are the same as in GR,
focusing first on the inspiral phase of coalescence.

In the inspiral phase of the binary's evolution it is appropriate to
use the PN approximation, an expansion in powers of
$v/c \sim (Gm/rc^2)^{1/2}$.  It is convenient to introduce a rescaled
version of the scalar field $\phi$: $\varphi \equiv \phi/\phi_0$,
where $\phi_0$ is the value of $\phi$ at infinity (assumed to be
constant).  Mirshekari and Will~\cite{Mirshekari:2013vb} found the
equations of motion for the bodies up to 2.5PN order.  Schematically,
the relative acceleration
$\mathbf{a} \equiv \mathbf{a}_1-\mathbf{a}_2$ takes the form
\begin{align}
a^i =& -\frac{G\alpha m}{r^2}\hat{n}^i+\frac{G\alpha m}{r^2}(A_\text{PN}\hat{n}^i+B_\text{PN}\dot{r}v^i)+\frac{8}{5}\eta\frac{(G\alpha m)^2}{r^3}(A_\text{1.5PN}\dot{r}\hat{n}^i-B_\text{1.5PN}v^i) \nonumber\\
&{}+\frac{G\alpha m}{r^2}(A_\text{2PN}\hat{n}^i+B_\text{2PN}\dot{r}v^i) \, ,
\end{align}
where $m \equiv m_1+m_2$, $\eta \equiv m_1m_2/m^2$, $r$ is the orbital
separation, $\mathbf{\hat{n}}$ is a unit vector pointing from body 2
to body 1, and $\mathbf{v} \equiv \mathbf{v}_1-\mathbf{v}_2$ is the
relative velocity.  The coefficients $A_\text{PN}$, $B_\text{PN}$,
$A_\text{1.5PN}$, $B_\text{1.5PN}$, $A_\text{2PN}$, and $B_\text{2PN}$
(which are typically time-dependent) are given
in~\cite{Mirshekari:2013vb}.  The symbol $G$ represents the
combination $(4+2\omega_0)/[\phi_0(3+2\omega_0)]$ [with
$\omega_0 \equiv \omega(\phi_0)$], which appears in the metric
component $g_{00}$ in the same manner as the gravitational constant
$G$ in GR.  However, the coupling in the Newtonian piece of the
equations of motion is not simply $G$ but $G\alpha$, where
\begin{equation}
\alpha \equiv \frac{3+2\omega_0}{4+2\omega_0}+\frac{(1-2s_1)(1-2s_2)}{4+2\omega_0} \,
\label{eq:alpha}
\end{equation}
and $s_i$ ($i=1\,,2$) are the sensitivities of the two objects:
\begin{equation}
s_A \equiv \left(\frac{d\ln M_{\scriptscriptstyle A}(\phi)}{d\ln \phi}\right)_{\phi=\phi_0} \,. \label{sensitivity}
\end{equation}
Higher-order derivatives of $M_{\scriptscriptstyle A}(\phi)$ are used
to define higher-order sensitivities, e.g.~$s'_{\scriptscriptstyle A}$
and $s''_{\scriptscriptstyle A}$.
Note that in GR radiation reaction begins at 2.5PN order (quadrupole
radiation), while in scalar-tensor gravity radiation reaction begins
at 1.5PN order, due to the presence of dipole radiation.

All deviations from GR can be characterized using a fairly small
number of parameters, all combinations of $\phi_0$, the Taylor
coefficients of $\omega(\phi)$, and the sensitivities $s_A$, $s_A'$,
and $s_A''$.  If one object in the system is a black hole (with the
other being a neutron star), the motion of the system is
indistinguishable from GR up to 1PN order.  All deviations beyond 1PN
order depend only on a single parameter, which is a function of
$\omega_0$ and the neutron star sensitivity.  Unfortunately, this
parameter alone (if measured) could not be used to distinguish between
Brans-Dicke theory and a more general scalar-tensor theory.

Going beyond the equations of motion, the next step is the calculation
of gravitational radiation.  The tensor part of the radiation, encoded
in $\tilde{h}^{ij}$, was computed up to 2PN order by
Lang~\cite{Lang:2013fna}.  All deviations depend on the same (small)
number of parameters that characterize the equations of motion.  For
black hole-neutron star systems, the waveform is indistinguishable from GR up to 1PN
order, and deviations at higher order depend only on the single
parameter described above.  For binary black hole systems, the waveform is {\em
  completely} indistinguishable from GR.  Scalar radiation has
recently been computed by Lang~\cite{Lang:2014osa} using a very
similar procedure. The dipole moment generates the lowest-order scalar
waves, which are of $-0.5$PN order:
\begin{equation}
\varphi = \frac{4G\mu\alpha^{1/2}}{R}\zeta\mathcal{S}_-(\mathbf{\hat{N}}\cdot \mathbf{v})
 \, ,
\label{eq:varphi}
\end{equation}
where $\mu \equiv m_1 m_2/m$ is the reduced mass, $\mathbf{\hat{N}}
\equiv \mathbf{x}/R$ is the direction from the source to the detector,
$\zeta \equiv 1/(4+2\omega_0)$, and
\begin{equation}
\mathcal{S}_- \equiv \alpha^{-1/2}(s_2-s_1) \, .
\end{equation}

Because computing the radiation up to 2PN order requires knowledge of
the monopole moment to 3PN order (relative to itself) and knowledge of
the dipole moment to 2.5PN order, Lang~\cite{Lang:2014osa} computed
the scalar waveform only to 1.5PN order.  The 1.5PN waveform is
described by the same set of parameters that describes the 2.5PN
equations of motion and the 2PN tensor waveform.  Again, the scalar
waveform vanishes for binary black hole systems (so that the GW signal is
indistinguishable from GR).

Lang~\cite{Lang:2014osa} used the tensor and scalar waveforms to
compute the total energy flux carried off to infinity to 1PN order.
A derivation of the quadrupole-order flux in tensor-multiscalar
theories (that agrees with Lang's results in the single-scalar limit)
can be found in~\cite{Damour:1992we}.
A similar calculation for compact binaries in the massive Brans-Dicke
theory was performed by Alsing et al.~\cite{Alsing:2011er} (see
also~\cite{Krause:1994ar,Perivolaropoulos:2009ak}).  In the notation
used above, and correcting a mistake in~\cite{Alsing:2011er}, they
found that the lowest-order flux is given by
\begin{equation}
\dot{E} = \frac{4}{3}\frac{\mu\eta}{r}\left(\frac{G\alpha m}{r}\right)^3\zeta\mathcal{S}_-^2\left[\frac{\omega^2-m_s^2}{\omega^2}\right]^{3/2}\Theta(\omega-m_s) \, ,
\label{eq:Alsingflux}
\end{equation}
where $\omega$ is the binary's orbital frequency, $m_s$ is the mass of
the scalar field, and $\Theta$ is the Heaviside function (i.e., in
massive Brans-Dicke theory, scalar dipole radiation is emitted only
when $\omega > m_s$).
%
% Alsing et al.~continued their calculation to 1PN order; however,
% those terms are incomplete and we do not list them here.

%\leo{Very minor changes here (see .tex)}
The emitted radiation has very special features for a binary black hole
system: from Eq.~\eqref{eq:alpha} and \eqref{eq:varphi} with
$s_1=s_2=1/2$ we see that the dominant terms are {\em identical} to
the equations of motion in GR, except for an unobservable mass
rescaling.
This result is a generalization to binary systems of
``no-scalar-hair'' theorems that apply to single
black holes~\cite{Hawking:1972qk}.  For generic mass ratio, Mirshekari and
Will proved this ``generalized no-hair theorem'' up to 2.5PN order,
but they conjectured that it should hold at all PN orders.  Indeed,
Ref.~\cite{Yunes:2011aa} has shown that the equations of motion are the same as
in GR at any PN order if one considers an extreme mass-ratio system
and works to lowest order in the mass ratio, and
the conjecture is also supported by numerical relativity
studies~\cite{Healy:2011ef,Berti:2013gfa}.  This ``generalized no-hair
theorem'' for binary black holes depends on some crucial assumptions:
vanishing scalar potential, asymptotically constant value of the
scalar field, and vanishing matter content.  If any one of these
assumptions breaks down, the black hole binary's behavior will differ from GR.

%% BibTeX users please use one of
%%\bibliographystyle{../TEMPLATE/spbasic}      % basic style, author-year citations
%%\bibliographystyle{../TEMPLATE/spmpsci}      % mathematics and physical sciences
%\bibliographystyle{../../TEMPLATE/spphys}       % APS-like style for physics
%%\bibliographystyle{apsrev4-1}
%\bibliography{../../TEMPLATE/master}   % name your BibTeX data base

\end{document}